\documentstyle[11pt]{article}
\newlength{\bredde}
\def\slash#1{\settowidth{\bredde}{$#1$}\ifmmode\,\raisebox{.15ex}{/}
\hspace*{-\bredde} #1\else$\,\raisebox{.15ex}{/}\hspace*{-\bredde} #1$\fi}
\newcommand{\beq}{\begin{equation}}
\newcommand{\eeq}{\end{equation}}

\newcommand{\lG}{\raise.3ex\hbox{$\stackrel{\leftarrow}{G}$}}
\newcommand{\lU}{\raise.3ex\hbox{$\stackrel{\leftarrow}{U}$}}
\newcommand{\lP}{\raise.3ex\hbox{$\stackrel{\leftarrow}{{\cal P}}$}}
\newcommand{\leta}{\raise.3ex\hbox{$\stackrel{\leftarrow}{\eta}$}}
\newcommand{\lOmega}{\raise.3ex\hbox{$\stackrel{\leftarrow}{\Omega}$}}
\newcommand{\ldr}{\raise.3ex\hbox{$\stackrel{\leftarrow}{\delta^r}$}}
\def\beqn{\begin{eqnarray}}
\def\eeqn{\end{eqnarray}}

\def\gtwid{\raise.3ex\hbox{$>$\kern-.75em\lower1ex\hbox{$\sim$}}}
\def\ltwid{\raise.3ex\hbox{$<$\kern-.75em\lower1ex\hbox{$\sim$}}}

\begin{document}
\topmargin -1.4cm
\oddsidemargin -0.8cm
\evensidemargin -0.8cm
\title{\Large{On Lattice QCD with Many Flavors}}

\vspace{0.9cm}

\author{
{\sc P.H. Damgaard$^a$, U.M. Heller$^b$, A. Krasnitz$^a$ and P. Olesen$^a$}
\\~\\~$^{a)}$ The Niels Bohr Institute\\Blegdamsvej 17\\DK-2100
Copenhagen\\Denmark\\~\\~$^{b)}$ SCRI, Florida State University\\ 
Tallahassee, FL 32306-4052\\USA
}
 
\maketitle
\vfill
\begin{abstract} 
We discuss the confining and chiral-symmetry breaking properties of
QCD with a large number of flavors $N_f$. In a Monte Carlo simulation
of QCD with $N_f =16$ staggered fermions, we find clear evidence of
a first order bulk phase transition which separates phases with broken
and unbroken chiral symmetry. This is consistent with extrapolations
of earlier studies with smaller $N_f$, and is also as expected from general
arguments. Thus, even when the perturbative 
renormalization group flow has a new infrared stable fixed point near 
the origin, lattice artifacts induce chiral symmetry breaking, and 
presumably confinement, at sufficiently strong coupling.
\end{abstract}
\vfill
\begin{flushleft}
NBI-HE-96-59 \\
FSU-SCRI-97-05\\
hep-lat/9701008
\end{flushleft}
\thispagestyle{empty}
\newpage

\setcounter{page}{1}
The confining properties of QCD are believed to depend sensitively on
the number of light quarks species $N_f$. For example, when $N_f \geq 17$
the theory is not asymptotically free in the ultraviolet, and is
presumably ``trivial'', with free quarks and gluons. Chiral symmetry
breaking is not expected to occur either. But an interesting change in
the phase structure may occur even before $N_f = 17$. It was noted
already in one of the first papers describing the two-loop beta function
of QCD that the two-loop term changes sign at a value of $N_f$ which
differs from that of the one-loop term \cite{caswell}. This means 
that even for $N_f < 17$ a new infrared fixed point can occur in the
renormalization group (RG) flow. From the two-loop beta function one obtains
for $SU(N)$, with $N_f$ flavors of quarks in the fundamental representation
 \cite{caswell,jones}
\beq
\frac{(g^*)^2}{16\pi^2}=\frac{11N-2N_f}{-34N^2+13NN_f-3N_f/N}
\label{twoloop}
\eeq
for the fixed point. For $SU(3)$ we thus get
\beq
\frac{(g^*)^2}{16\pi^2}=\frac{16.5-N_f}{-153+19N_f} ~,
\label{su3}
\eeq
where the right hand side of course should be positive. This leads to the
constraint $N_f\geq$ 8.053 (and $N_f \leq 16.5$). Taken at face value, 
this new fixed point thus occurs already for a number of flavors
$N^* = 9$. Of course, this prediction is perturbative,
and hence can only be trusted if the new fixed point 
self-consistently occurs close to the origin, which is certainly not the case
for $N_f=9$. On the other hand, it is
clear from these simple arguments that if we view the number of quark
flavors $N_f$ as a free, tunable (perhaps not even integer\footnote{This
is justified to any order of perturbation theory, since the Casimir operators
are polynomials in $N_f$.}) parameter,
then the RG flow of QCD will be substantially altered as $N_f$ is
increased. This happens {\em before} asymptotic freedom is lost. However, it is
clear that if $N_f=9$, corresponding to $(g^*)^2/16\pi^2\approx 0.417$, we are
so far from a perturbatively self-consistent region that the fixed point
has no a priori meaning, and it is extremely dangerous to attempt to make much
ado about $N_f=9$.

QCD with an infrared fixed point of the above kind was first analyzed in
some detail by Banks and Zaks \cite{Banks}, assuming that $N_f$ was large
enough to make the perturbative prediction valid. More recently, a study 
based on truncated Schwinger-Dyson equations has tried to determine more
accurately the ``critical number of flavors'' $N^*$ at which the 
large-distance dynamics of QCD is governed by an infrared fixed point 
\cite{Appelquist}. Another interesting approach, trying to go beyond the 
scheme-independent first two coefficients of the beta function and even
beyond the (known) third coefficient in some given scheme, is based on 
the large-$N_f$ expansion \cite{Gracey}. But it would obviously be
desirable to consider the whole issue in a more stringent non-perturbative
framework, in which the approximations can be controlled and improved in
a systematic manner. This suggests the use of lattice gauge theory.

We are therefore presently undertaking an investigation of the r\^{o}le 
played by $N_f$ in
lattice-regularized QCD, with $N_f$ (continuum) species of staggered
fermions, and the conventional Wilson gauge term 
($\beta \equiv 6/g^2$):
\beq
S_{gauge} = \beta \sum_p (1 - \frac{1}{3}{\mbox{\rm Re Tr}}\, U_p)
\label{action}
\eeq
where $U_p$ is the usual plaquette.

In this letter we shall report on some results for $N_f = 16$.
This is the largest integer value of $N_f$ which is compatible
with asymptotic freedom. On the other hand, it is also a value for
which the prediction of a new perturbative infrared fixed point most likely can
be trusted, since from eq. (\ref{twoloop}) one gets $(g^*)^2/16\pi^2 = 1/302$.
In principle, this fixed point could be singular, meaning that physical
quantities cannot be obtained by just putting $g=g^*$. However, in the
following we shall assume that perturbation theory is asymptotically
correct for small enough coupling as far as the first two (universal) terms
in the $\beta$-function are concerned. Then, for $g^*$ sufficiently small,
there should be no singular behaviour as the coupling $g$ approaches $g^*$.
Close to the origin, the RG flow is thus restricted to lie in the interval
$0 \leq g \leq g^*$, moving towards $g^*$ at large distances. From
general arguments we do not expect this theory to be confining.
What happens on the lattice? Consider the limit of infinitely strong
bare gauge coupling: $\beta = 0$. For any number of quark flavors
one can show that chiral symmetry is broken here 
\cite{chibreak1,chibreak2}. 
The effective Lagrangian, obtained by performing the gauge (link) 
integrations, can in this case explicitly be written in terms of
meson and baryon fields \cite{chibreak2}. 
No free quark lines are permitted by the
compact nature of the gauge integrations. It seems impossible not to
associate this limit of extremely strong coupling with a confining
phase. Since we have just argued from the other end of the coupling
constant line that the theory is unconfining near the conventional
continuum limit at $g = 0$, this means that a deconfining phase transition
must occur somewhere in between. This phase transition is {\em not}
expected to be associated with the infrared fixed point at $g = g^*$.
Rather, it should occur at much larger values of $g$, at the clash
between the extreme strong-coupling lattice artifacts and the
continuation beyond $g^*$ of the phase predicted by perturbation theory.

\vspace{0.3cm}

We now present some of our results.
We have performed simulations of the $N_f=16$ theory in the range of gauge 
couplings $4.0\leq\beta\leq 4.375$. For our
exploratory study of the phase structure we chose a relatively heavy staggered 
quark mass $m_q=0.1$. We then further explored the weak-coupling phase of the
theory by varying the quark mass between 0.025 and 0.15.
The R version \cite{ralg} of refreshed molecular
dynamics algorithm was used for sampling of gauge configurations, with momenta
refreshed every time unit. We used a microcanonical time step of $0.005$, 
a conservative choice for
which a convergence was observed in a multiflavor study by the Columbia group
\cite{Christ}. 

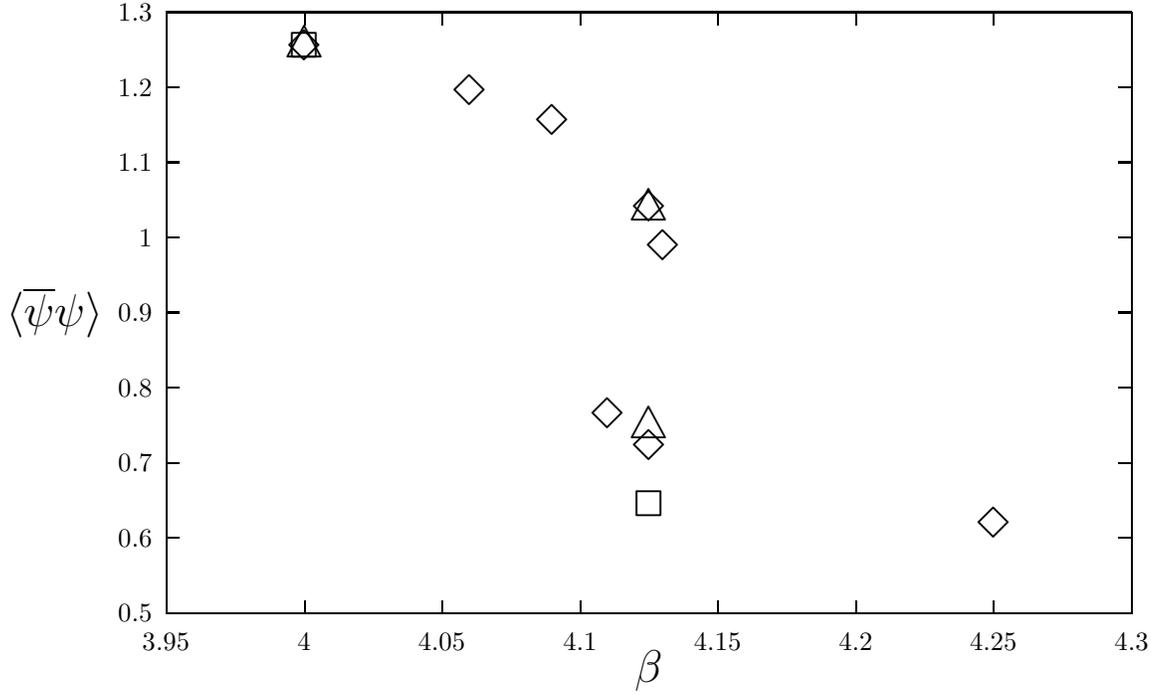
\begin{figure}
\begin{LARGE}
\setlength{\unitlength}{0.240900pt}
\ifx\plotpoint\undefined\newsavebox{\plotpoint}\fi
\sbox{\plotpoint}{\rule[-0.200pt]{0.400pt}{0.400pt}}%
\begin{picture}(1800,1080)(0,0)
\font\gnuplot=cmr10 at 10pt
\gnuplot
\sbox{\plotpoint}{\rule[-0.200pt]{0.400pt}{0.400pt}}%
\put(220.0,113.0){\rule[-0.200pt]{4.818pt}{0.400pt}}
\put(198,113){\makebox(0,0)[r]{0.5}}
\put(1716.0,113.0){\rule[-0.200pt]{4.818pt}{0.400pt}}
\put(220.0,231.0){\rule[-0.200pt]{4.818pt}{0.400pt}}
\put(198,231){\makebox(0,0)[r]{0.6}}
\put(1716.0,231.0){\rule[-0.200pt]{4.818pt}{0.400pt}}
\put(220.0,349.0){\rule[-0.200pt]{4.818pt}{0.400pt}}
\put(198,349){\makebox(0,0)[r]{0.7}}
\put(1716.0,349.0){\rule[-0.200pt]{4.818pt}{0.400pt}}
\put(220.0,467.0){\rule[-0.200pt]{4.818pt}{0.400pt}}
\put(198,467){\makebox(0,0)[r]{0.8}}
\put(1716.0,467.0){\rule[-0.200pt]{4.818pt}{0.400pt}}
\put(220.0,585.0){\rule[-0.200pt]{4.818pt}{0.400pt}}
\put(198,585){\makebox(0,0)[r]{0.9}}
\put(1716.0,585.0){\rule[-0.200pt]{4.818pt}{0.400pt}}
\put(220.0,703.0){\rule[-0.200pt]{4.818pt}{0.400pt}}
\put(198,703){\makebox(0,0)[r]{1}}
\put(1716.0,703.0){\rule[-0.200pt]{4.818pt}{0.400pt}}
\put(220.0,821.0){\rule[-0.200pt]{4.818pt}{0.400pt}}
\put(198,821){\makebox(0,0)[r]{1.1}}
\put(1716.0,821.0){\rule[-0.200pt]{4.818pt}{0.400pt}}
\put(220.0,939.0){\rule[-0.200pt]{4.818pt}{0.400pt}}
\put(198,939){\makebox(0,0)[r]{1.2}}
\put(1716.0,939.0){\rule[-0.200pt]{4.818pt}{0.400pt}}
\put(220.0,1057.0){\rule[-0.200pt]{4.818pt}{0.400pt}}
\put(198,1057){\makebox(0,0)[r]{1.3}}
\put(1716.0,1057.0){\rule[-0.200pt]{4.818pt}{0.400pt}}
\put(220.0,113.0){\rule[-0.200pt]{0.400pt}{4.818pt}}
\put(220,68){\makebox(0,0){3.95}}
\put(220.0,1037.0){\rule[-0.200pt]{0.400pt}{4.818pt}}
\put(437.0,113.0){\rule[-0.200pt]{0.400pt}{4.818pt}}
\put(437,68){\makebox(0,0){4}}
\put(437.0,1037.0){\rule[-0.200pt]{0.400pt}{4.818pt}}
\put(653.0,113.0){\rule[-0.200pt]{0.400pt}{4.818pt}}
\put(653,68){\makebox(0,0){4.05}}
\put(653.0,1037.0){\rule[-0.200pt]{0.400pt}{4.818pt}}
\put(870.0,113.0){\rule[-0.200pt]{0.400pt}{4.818pt}}
\put(870,68){\makebox(0,0){4.1}}
\put(870.0,1037.0){\rule[-0.200pt]{0.400pt}{4.818pt}}
\put(1086.0,113.0){\rule[-0.200pt]{0.400pt}{4.818pt}}
\put(1086,68){\makebox(0,0){4.15}}
\put(1086.0,1037.0){\rule[-0.200pt]{0.400pt}{4.818pt}}
\put(1303.0,113.0){\rule[-0.200pt]{0.400pt}{4.818pt}}
\put(1303,68){\makebox(0,0){4.2}}
\put(1303.0,1037.0){\rule[-0.200pt]{0.400pt}{4.818pt}}
\put(1519.0,113.0){\rule[-0.200pt]{0.400pt}{4.818pt}}
\put(1519,68){\makebox(0,0){4.25}}
\put(1519.0,1037.0){\rule[-0.200pt]{0.400pt}{4.818pt}}
\put(1736.0,113.0){\rule[-0.200pt]{0.400pt}{4.818pt}}
\put(1736,68){\makebox(0,0){4.3}}
\put(1736.0,1037.0){\rule[-0.200pt]{0.400pt}{4.818pt}}
\put(220.0,113.0){\rule[-0.200pt]{365.204pt}{0.400pt}}
\put(1736.0,113.0){\rule[-0.200pt]{0.400pt}{227.410pt}}
\put(220.0,1057.0){\rule[-0.200pt]{365.204pt}{0.400pt}}
\put(45,585){\makebox(0,0){$\langle{\overline\psi}\psi\rangle$}}
\put(978,23){\makebox(0,0){$\beta$}}
\put(220.0,113.0){\rule[-0.200pt]{0.400pt}{227.410pt}}
\put(437,1000){\raisebox{-.8pt}{\makebox(0,0){$\Diamond$}}}
\put(696,930){\raisebox{-.8pt}{\makebox(0,0){$\Diamond$}}}
\put(826,883){\raisebox{-.8pt}{\makebox(0,0){$\Diamond$}}}
\put(913,423){\raisebox{-.8pt}{\makebox(0,0){$\Diamond$}}}
\put(978,748){\raisebox{-.8pt}{\makebox(0,0){$\Diamond$}}}
\put(978,372){\raisebox{-.8pt}{\makebox(0,0){$\Diamond$}}}
\put(1000,686){\raisebox{-.8pt}{\makebox(0,0){$\Diamond$}}}
\put(1519,250){\raisebox{-.8pt}{\makebox(0,0){$\Diamond$}}}
\sbox{\plotpoint}{\rule[-0.400pt]{0.800pt}{0.800pt}}%
\put(437,1000){\raisebox{-.8pt}{\makebox(0,0){$\Box$}}}
\put(978,281){\raisebox{-.8pt}{\makebox(0,0){$\Box$}}}
\sbox{\plotpoint}{\rule[-0.500pt]{1.000pt}{1.000pt}}%
\sbox{\plotpoint}{\rule[-0.600pt]{1.200pt}{1.200pt}}%
\put(437,1004){\makebox(0,0){$\triangle$}}
\put(978,748){\makebox(0,0){$\triangle$}}
\put(978,406){\makebox(0,0){$\triangle$}}
\sbox{\plotpoint}{\rule[-0.500pt]{1.000pt}{1.000pt}}%
\end{picture}
\end{LARGE}
\caption{The $\beta$ dependence of the chiral order parameter for lattice
sizes $8^3\times 16$ (diamonds), $12^4$ (triangles),
and $6^3\times 16$ (squares). The quark mass is $am_q=0.1$. 
The error bars are smaller than the plotting
symbols.}
\label{pbp}
\end{figure}

We have measured, in addition to the chiral order parameter
$\langle{\overline\psi}\psi\rangle$ and a variety of Wilson loops and
correlation functions, the
gauge-invariant fermion correlation functions appearing in the  
Fredenhagen-Marcu (FM) order parameter \cite{fm} (see below). 
We find strong evidence that the system
undergoes a first order phase transition, with a metastability region ranging
at least between $\beta=4.11$ and $\beta=4.13$. To clarify: our criterion for
metastability is the existence of two phases at a given value of $\beta$, 
with each phase having a lifetime of at least 200 microcanonical time units. 
In that region there are clear discontinuities in a number of observables: 
the chiral order parameter, the average action, and the string tension. 
Most importantly, as Figures \ref{pbp} -- \ref{st33} show, the location and 
the strength of the transition do not show any significant sensitivity to 
variations in the lattice size (although observation of metastability is 
difficult for small sizes). What we are seeing 
is thus most likely a bulk, rather than a finite-temperature phase
transition. The nature of the transition is best revealed by the behavior 
of the chiral order parameter $\langle{\overline\psi}\psi\rangle$ and  
the string tension. The chiral
order parameter decreases sharply in going from the low-$\beta$ to the
high-$\beta$ phase, suggesting that genuine chiral symmetry
restoration occurs in the limit of vanishing quark mass (see below). 
Such restoration 
is usually also associated with a transition from confinement to 
deconfinement, but this is not {\em a priori} obvious. Of course, the
notion of confinement in the presence of dynamical quarks in the
fundamental representation of the gauge group is highly
non-trivial (although an order parameter to distinguish between
confinement and deconfinement in this case does exist, see below) due
to the presence of quark-antiquark screening. 
Nevertheless, we would expect that the ``string tension'' formally defined,
for intermediate distances, through a Creutz ratio of Wilson loops
\beq
\sigma(x,t)\equiv-{\rm ln}\left({{W(x,t)W(x-1,t-1)}\over{W(x,t-1)W(x-1,t)}}
\right) ~,\label{Creutz}
\eeq
should drop discontinuously at a first-order deconfining phase transition.
This is indeed what we observe in Figure \ref{st33}. The residual $\sigma$
on the relatively small coupling side ($\beta \geq 4.12$) can at least
qualitatively be understood as a ``Coulomb''-like term. The contribution of a
term of the form $e/R$ ($e$ is some effective coupling, here a free parameter)
to $\sigma(x,t)$ is $-e/x(x-1)$,
so for $x$=3 we have a contribution $-e/6\simeq 0.17$ if $e$ is of order
-1. The residual $\sigma$ on the relatively small coupling side ($\beta
\geq 4.12$) of Fig. \ref{st33} can thus easily be understood by some effective 
``Coulomb'' term, and not due to any string tension. 
A value $e\approx -1$ is certainly not an unreasonable order of magnitude
for a ``Coulomb''-coefficient in the strong coupling regime. For the much
smaller number of flavors $N_f = 2$, a very accurate fit yields a value
of $e \simeq -0.32$ at $\beta = 5.6$, while for $N_f = 0$ (quenched
simulations) the corresponding fit at the same value gives $e \simeq -0.29$
\cite{nfpot}. So the magnitude of this parameter increases with $N_f$ at 
a fixed $\beta$-value and is also expected to increase with the coupling $g$,
consistent with our present estimate. Note also that the sign of this
term is consistent with that of the $N_f = 2$ theory.

A confinement criterion valid even in the presence of dynamical fermions 
has been proposed by
Fredenhagen and Marcu \cite{fm}. We have tried to measure their 
corresponding order parameter 
\beq
\rho ~\equiv~ \lim_{R \to \infty} \frac{\langle\bar{\psi}(0)U(R,T)\psi(R)
\rangle}{\sqrt{W(R,2T)}} ~, ~~~~~~~~R/T {\mbox{\rm ~~ kept fixed}}
\label{FM}
\eeq
as well. Here the numerator is a ``staple'' constructed by parallel 
displacement along a path of length $T$ in the (arbitrarily chosen )``time'' 
direction. The numerator is (the square root of) a Wilson loop of size 
$R\times 2T$.
The small sizes of our lattices obviously did not allow us to explore 
large space and time separations, as required by the FM order parameter. 
Moreover, our configuration sample was too small and did not permit us to 
determine the FM order parameter for the larger distances we actually had
available. For smaller separations, 
where the measurement was possible, the finite-distance version of
order parameter (\ref{FM}) exhibited an somewhat unusual 
behavior. Namely, at zero separation, where the FM parameter coincides 
with $\langle{\overline\psi}\psi\rangle$, the broken-phase value is larger 
than the symmetric-phase one. However, the FM parameter decreases with 
separation slower in the symmetric phase than it does in the broken phase, 
until finally the relative magnitude between the two is reversed. 
The only way to reconcile our
data with the assumption, still unproven, that confinement implies
chiral symmetry breaking, is by saying that the FM parameter approaches 
zero in the deconfined phase {\em slower} than it approaches a constant 
non-vanishing value in the confined phase.
If so, considerably larger lattices are required for the FM parameter to
be useful as a confinement criterion for these couplings, and with this number
of $N_f$.

At strong coupling, confinement/deconfinement can be checked explicitly
by use of the FM order parameter. At infinitely strong coupling ($\beta = 0$)
one can compute its value by either a hopping
parameter expansion (in the case of Wilson fermions), or by an expansion
in $1/m$ (in the case of staggered fermions).  
At infinitely strong coupling one finds, as expected, that
the FM order parameter is nonvanishing (to leading order it is even
constant, and the large-$R$ limit is reached immediately), indicating 
confinement. This holds for any number of flavors $N_f$.

\begin{figure}
\begin{LARGE}
\setlength{\unitlength}{0.240900pt}
\ifx\plotpoint\undefined\newsavebox{\plotpoint}\fi
\sbox{\plotpoint}{\rule[-0.200pt]{0.400pt}{0.400pt}}%
\begin{picture}(1800,1080)(0,0)
\font\gnuplot=cmr10 at 10pt
\gnuplot
\sbox{\plotpoint}{\rule[-0.200pt]{0.400pt}{0.400pt}}%
\put(220.0,113.0){\rule[-0.200pt]{4.818pt}{0.400pt}}
\put(198,113){\makebox(0,0)[r]{-2.8}}
\put(1716.0,113.0){\rule[-0.200pt]{4.818pt}{0.400pt}}
\put(220.0,218.0){\rule[-0.200pt]{4.818pt}{0.400pt}}
\put(198,218){\makebox(0,0)[r]{-2.7}}
\put(1716.0,218.0){\rule[-0.200pt]{4.818pt}{0.400pt}}
\put(220.0,323.0){\rule[-0.200pt]{4.818pt}{0.400pt}}
\put(198,323){\makebox(0,0)[r]{-2.6}}
\put(1716.0,323.0){\rule[-0.200pt]{4.818pt}{0.400pt}}
\put(220.0,428.0){\rule[-0.200pt]{4.818pt}{0.400pt}}
\put(198,428){\makebox(0,0)[r]{-2.5}}
\put(1716.0,428.0){\rule[-0.200pt]{4.818pt}{0.400pt}}
\put(220.0,533.0){\rule[-0.200pt]{4.818pt}{0.400pt}}
\put(198,533){\makebox(0,0)[r]{-2.4}}
\put(1716.0,533.0){\rule[-0.200pt]{4.818pt}{0.400pt}}
\put(220.0,637.0){\rule[-0.200pt]{4.818pt}{0.400pt}}
\put(198,637){\makebox(0,0)[r]{-2.3}}
\put(1716.0,637.0){\rule[-0.200pt]{4.818pt}{0.400pt}}
\put(220.0,742.0){\rule[-0.200pt]{4.818pt}{0.400pt}}
\put(198,742){\makebox(0,0)[r]{-2.2}}
\put(1716.0,742.0){\rule[-0.200pt]{4.818pt}{0.400pt}}
\put(220.0,847.0){\rule[-0.200pt]{4.818pt}{0.400pt}}
\put(198,847){\makebox(0,0)[r]{-2.1}}
\put(1716.0,847.0){\rule[-0.200pt]{4.818pt}{0.400pt}}
\put(220.0,952.0){\rule[-0.200pt]{4.818pt}{0.400pt}}
\put(198,952){\makebox(0,0)[r]{-2}}
\put(1716.0,952.0){\rule[-0.200pt]{4.818pt}{0.400pt}}
\put(220.0,1057.0){\rule[-0.200pt]{4.818pt}{0.400pt}}
\put(198,1057){\makebox(0,0)[r]{-1.9}}
\put(1716.0,1057.0){\rule[-0.200pt]{4.818pt}{0.400pt}}
\put(220.0,113.0){\rule[-0.200pt]{0.400pt}{4.818pt}}
\put(220,68){\makebox(0,0){3.95}}
\put(220.0,1037.0){\rule[-0.200pt]{0.400pt}{4.818pt}}
\put(437.0,113.0){\rule[-0.200pt]{0.400pt}{4.818pt}}
\put(437,68){\makebox(0,0){4}}
\put(437.0,1037.0){\rule[-0.200pt]{0.400pt}{4.818pt}}
\put(653.0,113.0){\rule[-0.200pt]{0.400pt}{4.818pt}}
\put(653,68){\makebox(0,0){4.05}}
\put(653.0,1037.0){\rule[-0.200pt]{0.400pt}{4.818pt}}
\put(870.0,113.0){\rule[-0.200pt]{0.400pt}{4.818pt}}
\put(870,68){\makebox(0,0){4.1}}
\put(870.0,1037.0){\rule[-0.200pt]{0.400pt}{4.818pt}}
\put(1086.0,113.0){\rule[-0.200pt]{0.400pt}{4.818pt}}
\put(1086,68){\makebox(0,0){4.15}}
\put(1086.0,1037.0){\rule[-0.200pt]{0.400pt}{4.818pt}}
\put(1303.0,113.0){\rule[-0.200pt]{0.400pt}{4.818pt}}
\put(1303,68){\makebox(0,0){4.2}}
\put(1303.0,1037.0){\rule[-0.200pt]{0.400pt}{4.818pt}}
\put(1519.0,113.0){\rule[-0.200pt]{0.400pt}{4.818pt}}
\put(1519,68){\makebox(0,0){4.25}}
\put(1519.0,1037.0){\rule[-0.200pt]{0.400pt}{4.818pt}}
\put(1736.0,113.0){\rule[-0.200pt]{0.400pt}{4.818pt}}
\put(1736,68){\makebox(0,0){4.3}}
\put(1736.0,1037.0){\rule[-0.200pt]{0.400pt}{4.818pt}}
\put(220.0,113.0){\rule[-0.200pt]{365.204pt}{0.400pt}}
\put(1736.0,113.0){\rule[-0.200pt]{0.400pt}{227.410pt}}
\put(220.0,1057.0){\rule[-0.200pt]{365.204pt}{0.400pt}}
\put(45,585){\makebox(0,0){$\langle S/V \rangle$}}
\put(978,23){\makebox(0,0){$\beta$}}
\put(220.0,113.0){\rule[-0.200pt]{0.400pt}{227.410pt}}
\put(437,1008){\raisebox{-.8pt}{\makebox(0,0){$\Diamond$}}}
\put(696,890){\raisebox{-.8pt}{\makebox(0,0){$\Diamond$}}}
\put(826,826){\raisebox{-.8pt}{\makebox(0,0){$\Diamond$}}}
\put(913,409){\raisebox{-.8pt}{\makebox(0,0){$\Diamond$}}}
\put(978,682){\raisebox{-.8pt}{\makebox(0,0){$\Diamond$}}}
\put(978,362){\raisebox{-.8pt}{\makebox(0,0){$\Diamond$}}}
\put(1000,608){\raisebox{-.8pt}{\makebox(0,0){$\Diamond$}}}
\put(1519,208){\raisebox{-.8pt}{\makebox(0,0){$\Diamond$}}}
\sbox{\plotpoint}{\rule[-0.400pt]{0.800pt}{0.800pt}}%
\put(437,1007){\raisebox{-.8pt}{\makebox(0,0){$\Box$}}}
\put(978,314){\raisebox{-.8pt}{\makebox(0,0){$\Box$}}}
\sbox{\plotpoint}{\rule[-0.500pt]{1.000pt}{1.000pt}}%
\sbox{\plotpoint}{\rule[-0.600pt]{1.200pt}{1.200pt}}%
\put(437,1003){\makebox(0,0){$\triangle$}}
\put(978,664){\makebox(0,0){$\triangle$}}
\put(978,385){\makebox(0,0){$\triangle$}}
\end{picture}
\end{LARGE}
\caption{The $\beta$ dependence of the lattice action per site for lattice
sizes $8^3\times 16$ (diamonds), $12^4$ (triangles),
and $6^3\times 16$ (squares). The quark mass is $am_q=0.1$. 
The error bars are smaller than the plotting
symbols.}
\label{act}
\end{figure}
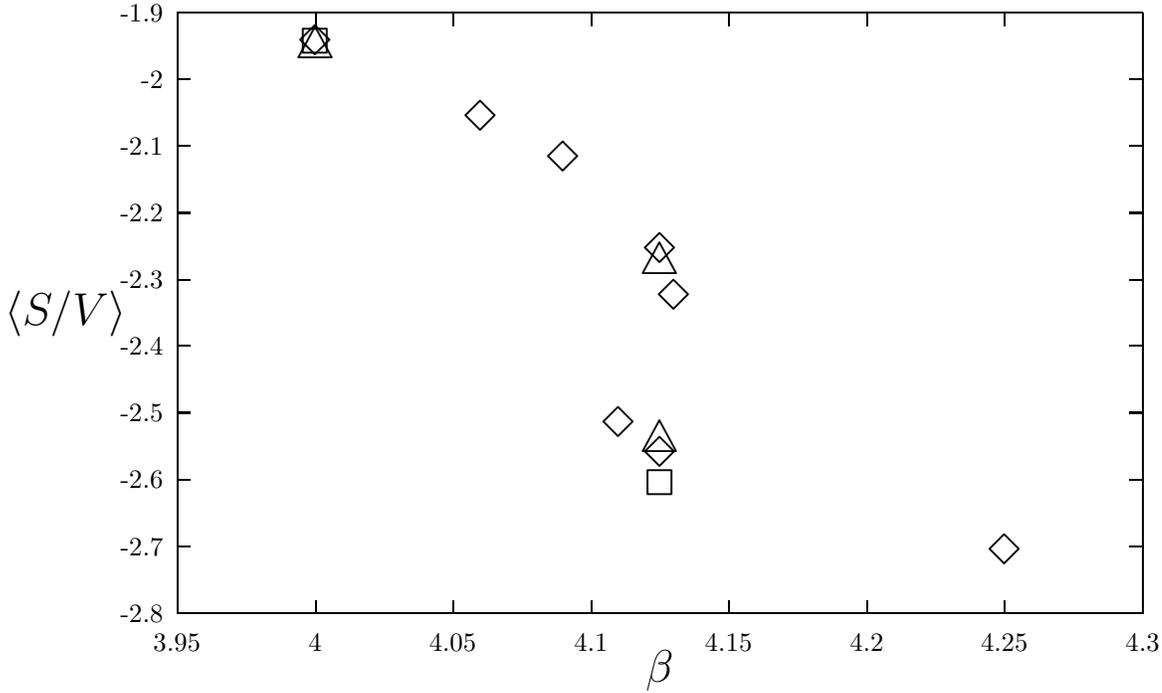
\begin{figure}
\begin{LARGE}
\setlength{\unitlength}{0.240900pt}
\ifx\plotpoint\undefined\newsavebox{\plotpoint}\fi
\sbox{\plotpoint}{\rule[-0.200pt]{0.400pt}{0.400pt}}%
\begin{picture}(1800,1080)(0,0)
\font\gnuplot=cmr10 at 10pt
\gnuplot
\sbox{\plotpoint}{\rule[-0.200pt]{0.400pt}{0.400pt}}%
\put(220.0,113.0){\rule[-0.200pt]{365.204pt}{0.400pt}}
\put(220.0,113.0){\rule[-0.200pt]{4.818pt}{0.400pt}}
\put(198,113){\makebox(0,0)[r]{0}}
\put(1716.0,113.0){\rule[-0.200pt]{4.818pt}{0.400pt}}
\put(220.0,270.0){\rule[-0.200pt]{4.818pt}{0.400pt}}
\put(198,270){\makebox(0,0)[r]{0.2}}
\put(1716.0,270.0){\rule[-0.200pt]{4.818pt}{0.400pt}}
\put(220.0,428.0){\rule[-0.200pt]{4.818pt}{0.400pt}}
\put(198,428){\makebox(0,0)[r]{0.4}}
\put(1716.0,428.0){\rule[-0.200pt]{4.818pt}{0.400pt}}
\put(220.0,585.0){\rule[-0.200pt]{4.818pt}{0.400pt}}
\put(198,585){\makebox(0,0)[r]{0.6}}
\put(1716.0,585.0){\rule[-0.200pt]{4.818pt}{0.400pt}}
\put(220.0,742.0){\rule[-0.200pt]{4.818pt}{0.400pt}}
\put(198,742){\makebox(0,0)[r]{0.8}}
\put(1716.0,742.0){\rule[-0.200pt]{4.818pt}{0.400pt}}
\put(220.0,900.0){\rule[-0.200pt]{4.818pt}{0.400pt}}
\put(198,900){\makebox(0,0)[r]{1}}
\put(1716.0,900.0){\rule[-0.200pt]{4.818pt}{0.400pt}}
\put(220.0,1057.0){\rule[-0.200pt]{4.818pt}{0.400pt}}
\put(198,1057){\makebox(0,0)[r]{1.2}}
\put(1716.0,1057.0){\rule[-0.200pt]{4.818pt}{0.400pt}}
\put(220.0,113.0){\rule[-0.200pt]{0.400pt}{4.818pt}}
\put(220,68){\makebox(0,0){3.95}}
\put(220.0,1037.0){\rule[-0.200pt]{0.400pt}{4.818pt}}
\put(380.0,113.0){\rule[-0.200pt]{0.400pt}{4.818pt}}
\put(380,68){\makebox(0,0){4}}
\put(380.0,1037.0){\rule[-0.200pt]{0.400pt}{4.818pt}}
\put(539.0,113.0){\rule[-0.200pt]{0.400pt}{4.818pt}}
\put(539,68){\makebox(0,0){4.05}}
\put(539.0,1037.0){\rule[-0.200pt]{0.400pt}{4.818pt}}
\put(699.0,113.0){\rule[-0.200pt]{0.400pt}{4.818pt}}
\put(699,68){\makebox(0,0){4.1}}
\put(699.0,1037.0){\rule[-0.200pt]{0.400pt}{4.818pt}}
\put(858.0,113.0){\rule[-0.200pt]{0.400pt}{4.818pt}}
\put(858,68){\makebox(0,0){4.15}}
\put(858.0,1037.0){\rule[-0.200pt]{0.400pt}{4.818pt}}
\put(1018.0,113.0){\rule[-0.200pt]{0.400pt}{4.818pt}}
\put(1018,68){\makebox(0,0){4.2}}
\put(1018.0,1037.0){\rule[-0.200pt]{0.400pt}{4.818pt}}
\put(1177.0,113.0){\rule[-0.200pt]{0.400pt}{4.818pt}}
\put(1177,68){\makebox(0,0){4.25}}
\put(1177.0,1037.0){\rule[-0.200pt]{0.400pt}{4.818pt}}
\put(1337.0,113.0){\rule[-0.200pt]{0.400pt}{4.818pt}}
\put(1337,68){\makebox(0,0){4.3}}
\put(1337.0,1037.0){\rule[-0.200pt]{0.400pt}{4.818pt}}
\put(1497.0,113.0){\rule[-0.200pt]{0.400pt}{4.818pt}}
\put(1497,68){\makebox(0,0){4.35}}
\put(1497.0,1037.0){\rule[-0.200pt]{0.400pt}{4.818pt}}
\put(1656.0,113.0){\rule[-0.200pt]{0.400pt}{4.818pt}}
\put(1656,68){\makebox(0,0){4.4}}
\put(1656.0,1037.0){\rule[-0.200pt]{0.400pt}{4.818pt}}
\put(220.0,113.0){\rule[-0.200pt]{365.204pt}{0.400pt}}
\put(1736.0,113.0){\rule[-0.200pt]{0.400pt}{227.410pt}}
\put(220.0,1057.0){\rule[-0.200pt]{365.204pt}{0.400pt}}
\put(45,585){\makebox(0,0){$\sigma$}}
\put(978,23){\makebox(0,0){$\beta$}}
\put(220.0,113.0){\rule[-0.200pt]{0.400pt}{227.410pt}}
\put(380,760){\raisebox{-.8pt}{\makebox(0,0){$\Diamond$}}}
\put(571,550){\raisebox{-.8pt}{\makebox(0,0){$\Diamond$}}}
\put(667,555){\raisebox{-.8pt}{\makebox(0,0){$\Diamond$}}}
\put(731,276){\raisebox{-.8pt}{\makebox(0,0){$\Diamond$}}}
\put(779,434){\raisebox{-.8pt}{\makebox(0,0){$\Diamond$}}}
\put(779,252){\raisebox{-.8pt}{\makebox(0,0){$\Diamond$}}}
\put(794,394){\raisebox{-.8pt}{\makebox(0,0){$\Diamond$}}}
\put(1177,213){\raisebox{-.8pt}{\makebox(0,0){$\Diamond$}}}
\put(1576,199){\raisebox{-.8pt}{\makebox(0,0){$\Diamond$}}}
\put(380.0,540.0){\rule[-0.200pt]{0.400pt}{105.996pt}}
\put(370.0,540.0){\rule[-0.200pt]{4.818pt}{0.400pt}}
\put(370.0,980.0){\rule[-0.200pt]{4.818pt}{0.400pt}}
\put(571.0,443.0){\rule[-0.200pt]{0.400pt}{51.553pt}}
\put(561.0,443.0){\rule[-0.200pt]{4.818pt}{0.400pt}}
\put(561.0,657.0){\rule[-0.200pt]{4.818pt}{0.400pt}}
\put(667.0,475.0){\rule[-0.200pt]{0.400pt}{38.303pt}}
\put(657.0,475.0){\rule[-0.200pt]{4.818pt}{0.400pt}}
\put(657.0,634.0){\rule[-0.200pt]{4.818pt}{0.400pt}}
\put(731.0,248.0){\rule[-0.200pt]{0.400pt}{13.490pt}}
\put(721.0,248.0){\rule[-0.200pt]{4.818pt}{0.400pt}}
\put(721.0,304.0){\rule[-0.200pt]{4.818pt}{0.400pt}}
\put(779.0,392.0){\rule[-0.200pt]{0.400pt}{19.995pt}}
\put(769.0,392.0){\rule[-0.200pt]{4.818pt}{0.400pt}}
\put(769.0,475.0){\rule[-0.200pt]{4.818pt}{0.400pt}}
\put(779.0,238.0){\rule[-0.200pt]{0.400pt}{6.986pt}}
\put(769.0,238.0){\rule[-0.200pt]{4.818pt}{0.400pt}}
\put(769.0,267.0){\rule[-0.200pt]{4.818pt}{0.400pt}}
\put(794.0,349.0){\rule[-0.200pt]{0.400pt}{21.440pt}}
\put(784.0,349.0){\rule[-0.200pt]{4.818pt}{0.400pt}}
\put(784.0,438.0){\rule[-0.200pt]{4.818pt}{0.400pt}}
\put(1177.0,193.0){\rule[-0.200pt]{0.400pt}{9.395pt}}
\put(1167.0,193.0){\rule[-0.200pt]{4.818pt}{0.400pt}}
\put(1167.0,232.0){\rule[-0.200pt]{4.818pt}{0.400pt}}
\put(1576.0,193.0){\rule[-0.200pt]{0.400pt}{2.891pt}}
\put(1566.0,193.0){\rule[-0.200pt]{4.818pt}{0.400pt}}
\put(1566.0,205.0){\rule[-0.200pt]{4.818pt}{0.400pt}}
\sbox{\plotpoint}{\rule[-0.400pt]{0.800pt}{0.800pt}}%
\put(380,617){\raisebox{-.8pt}{\makebox(0,0){$\Box$}}}
\put(779,212){\raisebox{-.8pt}{\makebox(0,0){$\Box$}}}
\put(380.0,394.0){\rule[-0.400pt]{0.800pt}{107.200pt}}
\put(370.0,394.0){\rule[-0.400pt]{4.818pt}{0.800pt}}
\put(370.0,839.0){\rule[-0.400pt]{4.818pt}{0.800pt}}
\put(779.0,197.0){\rule[-0.400pt]{0.800pt}{7.227pt}}
\put(769.0,197.0){\rule[-0.400pt]{4.818pt}{0.800pt}}
\put(769.0,227.0){\rule[-0.400pt]{4.818pt}{0.800pt}}
\sbox{\plotpoint}{\rule[-0.500pt]{1.000pt}{1.000pt}}%
\sbox{\plotpoint}{\rule[-0.600pt]{1.200pt}{1.200pt}}%
\put(380,452){\makebox(0,0){$\triangle$}}
\put(779,418){\makebox(0,0){$\triangle$}}
\put(779,265){\makebox(0,0){$\triangle$}}
\put(380.0,341.0){\rule[-0.600pt]{1.200pt}{53.480pt}}
\put(370.0,341.0){\rule[-0.600pt]{4.818pt}{1.200pt}}
\put(370.0,563.0){\rule[-0.600pt]{4.818pt}{1.200pt}}
\put(779.0,373.0){\rule[-0.600pt]{1.200pt}{21.922pt}}
\put(769.0,373.0){\rule[-0.600pt]{4.818pt}{1.200pt}}
\put(769.0,464.0){\rule[-0.600pt]{4.818pt}{1.200pt}}
\put(779.0,246.0){\rule[-0.600pt]{1.200pt}{9.154pt}}
\put(769.0,246.0){\rule[-0.600pt]{4.818pt}{1.200pt}}
\put(769.0,284.0){\rule[-0.600pt]{4.818pt}{1.200pt}}
\end{picture}
\end{LARGE}
\caption{The $\beta$ dependence of the string tension defined through a Creutz 
ratio $-\left(W(3,3)W(2,2)\right)/\left(W(3,2)W(2,3)\right)$ for lattice
sizes $8^3\times 16$ (diamonds), $12^4$ (triangles),
and $6^3\times 16$ (squares).}
\label{st33}
\end{figure}
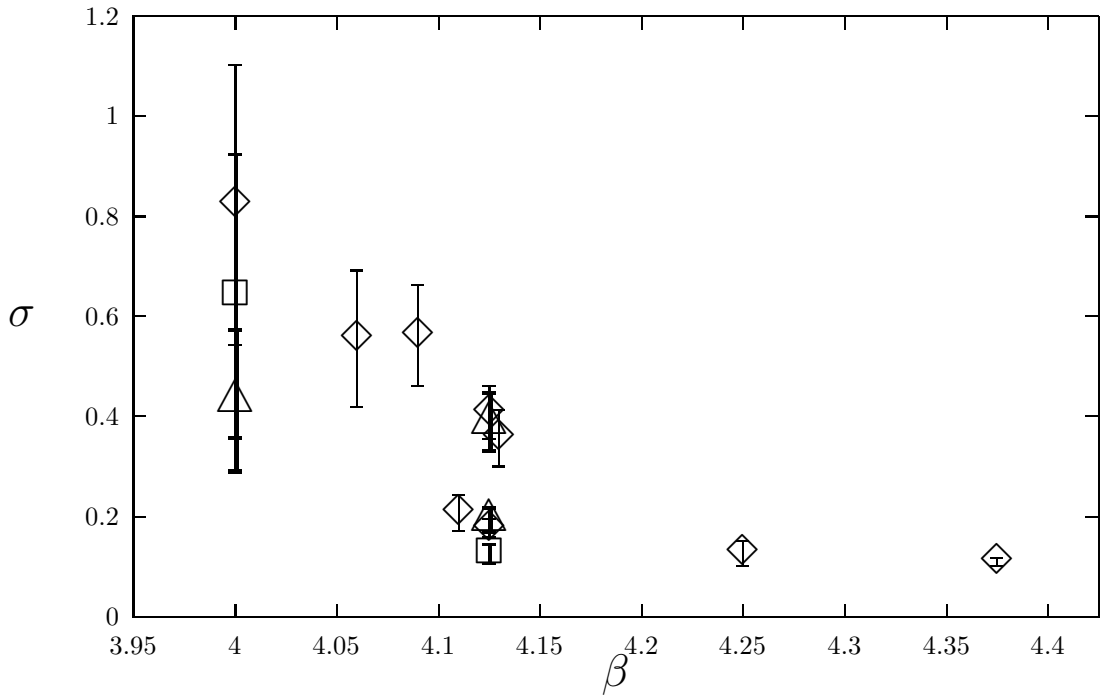
\vspace{1cm}

It is interesting to compare these results with earlier numerical work
for smaller values of $N_f$, and, since the phase transition 
described here occurs at quite strong coupling, with existing strong
coupling expansions. On the numerical side, zero-temperature
chiral phase transitions for moderately large values of $N_f$ 
were observed in refs. \cite{Kogut,Gavai}. For gauge group
$SU(3)$ and the same lattice action, they found a first-order phase
transition at $\beta_c = 4.47 \pm 0.04$ for $N_f = 12$ and lattice
sizes $6^4$ and $8^4$ \cite{Kogut}. For the same gauge group, the
first-order phase transition persisted for $N_f = 8$, and now occured at
$\beta_c \simeq 4.85$ on both $6^4$ and $8^4$ lattices. More
recently, in a very careful study, the Columbia
group found a number of intriguing and puzzling features around the
apparently bulk phase transition of the $N_f = 8$ theory
\cite{Christ}. Their value of the critical coupling equals
$\beta_c \simeq 4.73$ for lattices of sizes $16^3\times 8$ and
$16^4$, at least in rough agreement with the above number. We note also
the expected trend in the numbers: as $N_f$ is increased, the phase
transition moves toward stronger coupling. The fact that the phase
transition is pushed into strong coupling as $N_f$ grows, suggests 
that it eventually can be completely understood in terms of the strong 
coupling expansion.

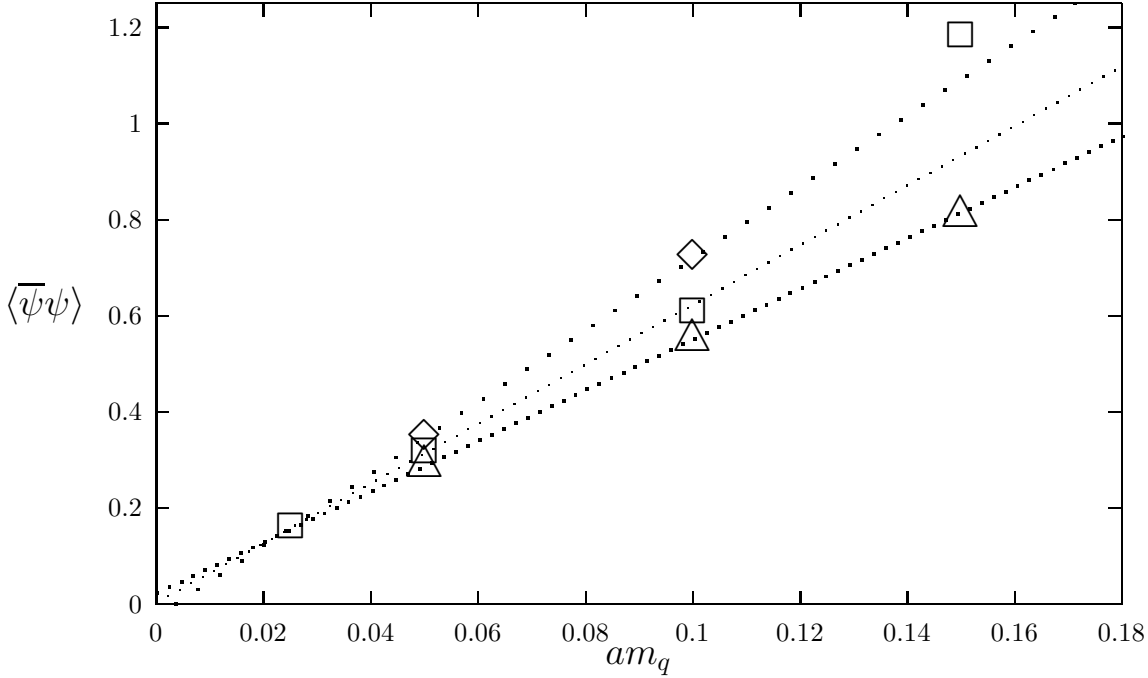
\begin{figure}
\begin{LARGE}
\setlength{\unitlength}{0.240900pt}
\ifx\plotpoint\undefined\newsavebox{\plotpoint}\fi
\sbox{\plotpoint}{\rule[-0.200pt]{0.400pt}{0.400pt}}%
\begin{picture}(1800,1080)(0,0)
\font\gnuplot=cmr10 at 10pt
\gnuplot
\sbox{\plotpoint}{\rule[-0.200pt]{0.400pt}{0.400pt}}%
\put(220.0,113.0){\rule[-0.200pt]{365.204pt}{0.400pt}}
\put(220.0,113.0){\rule[-0.200pt]{0.400pt}{227.410pt}}
\put(220.0,113.0){\rule[-0.200pt]{4.818pt}{0.400pt}}
\put(198,113){\makebox(0,0)[r]{0}}
\put(1716.0,113.0){\rule[-0.200pt]{4.818pt}{0.400pt}}
\put(220.0,264.0){\rule[-0.200pt]{4.818pt}{0.400pt}}
\put(198,264){\makebox(0,0)[r]{0.2}}
\put(1716.0,264.0){\rule[-0.200pt]{4.818pt}{0.400pt}}
\put(220.0,415.0){\rule[-0.200pt]{4.818pt}{0.400pt}}
\put(198,415){\makebox(0,0)[r]{0.4}}
\put(1716.0,415.0){\rule[-0.200pt]{4.818pt}{0.400pt}}
\put(220.0,566.0){\rule[-0.200pt]{4.818pt}{0.400pt}}
\put(198,566){\makebox(0,0)[r]{0.6}}
\put(1716.0,566.0){\rule[-0.200pt]{4.818pt}{0.400pt}}
\put(220.0,717.0){\rule[-0.200pt]{4.818pt}{0.400pt}}
\put(198,717){\makebox(0,0)[r]{0.8}}
\put(1716.0,717.0){\rule[-0.200pt]{4.818pt}{0.400pt}}
\put(220.0,868.0){\rule[-0.200pt]{4.818pt}{0.400pt}}
\put(198,868){\makebox(0,0)[r]{1}}
\put(1716.0,868.0){\rule[-0.200pt]{4.818pt}{0.400pt}}
\put(220.0,1019.0){\rule[-0.200pt]{4.818pt}{0.400pt}}
\put(198,1019){\makebox(0,0)[r]{1.2}}
\put(1716.0,1019.0){\rule[-0.200pt]{4.818pt}{0.400pt}}
\put(220.0,113.0){\rule[-0.200pt]{0.400pt}{4.818pt}}
\put(220,68){\makebox(0,0){0}}
\put(220.0,1037.0){\rule[-0.200pt]{0.400pt}{4.818pt}}
\put(388.0,113.0){\rule[-0.200pt]{0.400pt}{4.818pt}}
\put(388,68){\makebox(0,0){0.02}}
\put(388.0,1037.0){\rule[-0.200pt]{0.400pt}{4.818pt}}
\put(557.0,113.0){\rule[-0.200pt]{0.400pt}{4.818pt}}
\put(557,68){\makebox(0,0){0.04}}
\put(557.0,1037.0){\rule[-0.200pt]{0.400pt}{4.818pt}}
\put(725.0,113.0){\rule[-0.200pt]{0.400pt}{4.818pt}}
\put(725,68){\makebox(0,0){0.06}}
\put(725.0,1037.0){\rule[-0.200pt]{0.400pt}{4.818pt}}
\put(894.0,113.0){\rule[-0.200pt]{0.400pt}{4.818pt}}
\put(894,68){\makebox(0,0){0.08}}
\put(894.0,1037.0){\rule[-0.200pt]{0.400pt}{4.818pt}}
\put(1062.0,113.0){\rule[-0.200pt]{0.400pt}{4.818pt}}
\put(1062,68){\makebox(0,0){0.1}}
\put(1062.0,1037.0){\rule[-0.200pt]{0.400pt}{4.818pt}}
\put(1231.0,113.0){\rule[-0.200pt]{0.400pt}{4.818pt}}
\put(1231,68){\makebox(0,0){0.12}}
\put(1231.0,1037.0){\rule[-0.200pt]{0.400pt}{4.818pt}}
\put(1399.0,113.0){\rule[-0.200pt]{0.400pt}{4.818pt}}
\put(1399,68){\makebox(0,0){0.14}}
\put(1399.0,1037.0){\rule[-0.200pt]{0.400pt}{4.818pt}}
\put(1568.0,113.0){\rule[-0.200pt]{0.400pt}{4.818pt}}
\put(1568,68){\makebox(0,0){0.16}}
\put(1568.0,1037.0){\rule[-0.200pt]{0.400pt}{4.818pt}}
\put(1736.0,113.0){\rule[-0.200pt]{0.400pt}{4.818pt}}
\put(1736,68){\makebox(0,0){0.18}}
\put(1736.0,1037.0){\rule[-0.200pt]{0.400pt}{4.818pt}}
\put(220.0,113.0){\rule[-0.200pt]{365.204pt}{0.400pt}}
\put(1736.0,113.0){\rule[-0.200pt]{0.400pt}{227.410pt}}
\put(220.0,1057.0){\rule[-0.200pt]{365.204pt}{0.400pt}}
\put(45,585){\makebox(0,0){{\Large $\langle{\overline\psi}\psi\rangle$}}}
\put(978,23){\makebox(0,0){{\Large $am_q$}}}
\put(220.0,113.0){\rule[-0.200pt]{0.400pt}{227.410pt}}
\put(641,375){\raisebox{-.8pt}{\makebox(0,0){$\Diamond$}}}
\put(1062,657){\raisebox{-.8pt}{\makebox(0,0){$\Diamond$}}}
\sbox{\plotpoint}{\rule[-0.400pt]{0.800pt}{0.800pt}}%
\put(431,232){\raisebox{-.8pt}{\makebox(0,0){$\Box$}}}
\put(641,349){\raisebox{-.8pt}{\makebox(0,0){$\Box$}}}
\put(1062,570){\raisebox{-.8pt}{\makebox(0,0){$\Box$}}}
\put(1483,1003){\raisebox{-.8pt}{\makebox(0,0){$\Box$}}}
\sbox{\plotpoint}{\rule[-0.500pt]{1.000pt}{1.000pt}}%
\sbox{\plotpoint}{\rule[-0.600pt]{1.200pt}{1.200pt}}%
\put(641,330){\makebox(0,0){$\triangle$}}
\put(1062,529){\makebox(0,0){$\triangle$}}
\put(1483,723){\makebox(0,0){$\triangle$}}
\sbox{\plotpoint}{\rule[-0.500pt]{1.000pt}{1.000pt}}%
\put(249.00,113.00){\usebox{\plotpoint}}
\multiput(251,114)(34.539,23.026){0}{\usebox{\plotpoint}}
\multiput(266,124)(34.539,23.026){0}{\usebox{\plotpoint}}
\put(283.73,135.71){\usebox{\plotpoint}}
\multiput(297,144)(33.475,24.548){0}{\usebox{\plotpoint}}
\put(318.04,159.03){\usebox{\plotpoint}}
\multiput(327,165)(35.201,22.001){0}{\usebox{\plotpoint}}
\put(352.88,181.59){\usebox{\plotpoint}}
\multiput(358,185)(33.475,24.548){0}{\usebox{\plotpoint}}
\put(386.94,205.30){\usebox{\plotpoint}}
\multiput(388,206)(35.201,22.001){0}{\usebox{\plotpoint}}
\multiput(404,216)(34.539,23.026){0}{\usebox{\plotpoint}}
\put(421.70,227.98){\usebox{\plotpoint}}
\multiput(434,237)(35.201,22.001){0}{\usebox{\plotpoint}}
\put(456.15,251.10){\usebox{\plotpoint}}
\multiput(465,257)(34.539,23.026){0}{\usebox{\plotpoint}}
\put(490.58,274.28){\usebox{\plotpoint}}
\multiput(496,278)(34.539,23.026){0}{\usebox{\plotpoint}}
\put(525.07,297.38){\usebox{\plotpoint}}
\multiput(526,298)(35.201,22.001){0}{\usebox{\plotpoint}}
\multiput(542,308)(33.475,24.548){0}{\usebox{\plotpoint}}
\put(559.43,320.62){\usebox{\plotpoint}}
\multiput(572,329)(35.201,22.001){0}{\usebox{\plotpoint}}
\put(594.27,343.18){\usebox{\plotpoint}}
\multiput(603,349)(33.475,24.548){0}{\usebox{\plotpoint}}
\put(628.34,366.89){\usebox{\plotpoint}}
\multiput(633,370)(35.201,22.001){0}{\usebox{\plotpoint}}
\put(663.18,389.45){\usebox{\plotpoint}}
\multiput(664,390)(33.475,24.548){0}{\usebox{\plotpoint}}
\multiput(679,401)(35.201,22.001){0}{\usebox{\plotpoint}}
\put(697.54,412.69){\usebox{\plotpoint}}
\multiput(710,421)(34.539,23.026){0}{\usebox{\plotpoint}}
\put(732.01,435.82){\usebox{\plotpoint}}
\multiput(741,442)(34.539,23.026){0}{\usebox{\plotpoint}}
\put(766.46,458.97){\usebox{\plotpoint}}
\multiput(771,462)(35.201,22.001){0}{\usebox{\plotpoint}}
\put(800.86,482.17){\usebox{\plotpoint}}
\multiput(802,483)(34.539,23.026){0}{\usebox{\plotpoint}}
\multiput(817,493)(35.201,22.001){0}{\usebox{\plotpoint}}
\put(835.67,504.78){\usebox{\plotpoint}}
\multiput(848,513)(34.539,23.026){0}{\usebox{\plotpoint}}
\put(869.98,528.12){\usebox{\plotpoint}}
\multiput(878,534)(35.201,22.001){0}{\usebox{\plotpoint}}
\put(904.57,551.05){\usebox{\plotpoint}}
\multiput(909,554)(34.539,23.026){0}{\usebox{\plotpoint}}
\put(938.96,574.29){\usebox{\plotpoint}}
\multiput(940,575)(34.539,23.026){0}{\usebox{\plotpoint}}
\multiput(955,585)(34.539,23.026){0}{\usebox{\plotpoint}}
\put(973.56,597.22){\usebox{\plotpoint}}
\multiput(986,605)(33.475,24.548){0}{\usebox{\plotpoint}}
\put(1007.85,620.57){\usebox{\plotpoint}}
\multiput(1016,626)(35.201,22.001){0}{\usebox{\plotpoint}}
\put(1042.69,643.13){\usebox{\plotpoint}}
\multiput(1047,646)(33.475,24.548){0}{\usebox{\plotpoint}}
\put(1077.04,666.40){\usebox{\plotpoint}}
\multiput(1078,667)(34.539,23.026){0}{\usebox{\plotpoint}}
\multiput(1093,677)(34.539,23.026){0}{\usebox{\plotpoint}}
\put(1111.49,689.56){\usebox{\plotpoint}}
\multiput(1123,698)(35.201,22.001){0}{\usebox{\plotpoint}}
\put(1145.96,712.64){\usebox{\plotpoint}}
\multiput(1154,718)(34.539,23.026){0}{\usebox{\plotpoint}}
\put(1180.39,735.83){\usebox{\plotpoint}}
\multiput(1185,739)(34.539,23.026){0}{\usebox{\plotpoint}}
\put(1214.88,758.92){\usebox{\plotpoint}}
\multiput(1215,759)(35.201,22.001){0}{\usebox{\plotpoint}}
\multiput(1231,769)(33.475,24.548){0}{\usebox{\plotpoint}}
\put(1249.25,782.16){\usebox{\plotpoint}}
\multiput(1261,790)(35.201,22.001){0}{\usebox{\plotpoint}}
\put(1284.09,804.72){\usebox{\plotpoint}}
\multiput(1292,810)(33.475,24.548){0}{\usebox{\plotpoint}}
\put(1318.36,828.10){\usebox{\plotpoint}}
\multiput(1323,831)(34.539,23.026){0}{\usebox{\plotpoint}}
\put(1352.99,850.99){\usebox{\plotpoint}}
\multiput(1353,851)(34.539,23.026){0}{\usebox{\plotpoint}}
\multiput(1368,861)(34.207,23.517){0}{\usebox{\plotpoint}}
\put(1387.37,874.25){\usebox{\plotpoint}}
\multiput(1399,882)(34.539,23.026){0}{\usebox{\plotpoint}}
\put(1422.06,897.04){\usebox{\plotpoint}}
\multiput(1430,902)(33.475,24.548){0}{\usebox{\plotpoint}}
\put(1456.27,920.52){\usebox{\plotpoint}}
\multiput(1460,923)(35.201,22.001){0}{\usebox{\plotpoint}}
\multiput(1476,933)(34.539,23.026){0}{\usebox{\plotpoint}}
\put(1491.11,943.08){\usebox{\plotpoint}}
\multiput(1506,954)(35.201,22.001){0}{\usebox{\plotpoint}}
\put(1525.48,966.32){\usebox{\plotpoint}}
\multiput(1537,974)(34.539,23.026){0}{\usebox{\plotpoint}}
\put(1559.94,989.46){\usebox{\plotpoint}}
\multiput(1568,995)(34.539,23.026){0}{\usebox{\plotpoint}}
\put(1594.40,1012.60){\usebox{\plotpoint}}
\multiput(1598,1015)(34.539,23.026){0}{\usebox{\plotpoint}}
\put(1628.79,1035.85){\usebox{\plotpoint}}
\multiput(1629,1036)(34.539,23.026){0}{\usebox{\plotpoint}}
\multiput(1644,1046)(34.539,23.026){0}{\usebox{\plotpoint}}
\multiput(1659,1056)(37.129,18.564){0}{\usebox{\plotpoint}}
\put(1661,1057){\usebox{\plotpoint}}
\sbox{\plotpoint}{\rule[-0.200pt]{0.400pt}{0.400pt}}%
\put(220,115){\usebox{\plotpoint}}
\put(220.00,115.00){\usebox{\plotpoint}}
\put(237.92,125.46){\usebox{\plotpoint}}
\put(256.26,135.15){\usebox{\plotpoint}}
\put(274.29,145.42){\usebox{\plotpoint}}
\put(292.46,155.45){\usebox{\plotpoint}}
\put(310.72,165.32){\usebox{\plotpoint}}
\multiput(312,166)(17.798,10.679){0}{\usebox{\plotpoint}}
\put(328.62,175.81){\usebox{\plotpoint}}
\put(347.01,185.41){\usebox{\plotpoint}}
\put(365.00,195.74){\usebox{\plotpoint}}
\put(383.03,206.02){\usebox{\plotpoint}}
\put(401.38,215.69){\usebox{\plotpoint}}
\multiput(404,217)(17.798,10.679){0}{\usebox{\plotpoint}}
\put(419.28,226.17){\usebox{\plotpoint}}
\put(437.21,236.61){\usebox{\plotpoint}}
\put(455.54,246.32){\usebox{\plotpoint}}
\put(473.58,256.58){\usebox{\plotpoint}}
\put(491.75,266.61){\usebox{\plotpoint}}
\put(510.01,276.47){\usebox{\plotpoint}}
\multiput(511,277)(17.798,10.679){0}{\usebox{\plotpoint}}
\put(527.91,286.96){\usebox{\plotpoint}}
\put(546.29,296.58){\usebox{\plotpoint}}
\put(564.30,306.89){\usebox{\plotpoint}}
\put(582.48,316.89){\usebox{\plotpoint}}
\put(600.72,326.79){\usebox{\plotpoint}}
\multiput(603,328)(17.798,10.679){0}{\usebox{\plotpoint}}
\put(618.60,337.32){\usebox{\plotpoint}}
\put(636.87,347.18){\usebox{\plotpoint}}
\put(655.03,357.22){\usebox{\plotpoint}}
\put(673.08,367.45){\usebox{\plotpoint}}
\put(691.39,377.20){\usebox{\plotpoint}}
\put(709.34,387.60){\usebox{\plotpoint}}
\multiput(710,388)(18.314,9.767){0}{\usebox{\plotpoint}}
\put(727.60,397.46){\usebox{\plotpoint}}
\put(745.75,407.53){\usebox{\plotpoint}}
\put(763.84,417.70){\usebox{\plotpoint}}
\put(782.09,427.55){\usebox{\plotpoint}}
\put(800.09,437.86){\usebox{\plotpoint}}
\multiput(802,439)(18.314,9.767){0}{\usebox{\plotpoint}}
\put(818.33,447.75){\usebox{\plotpoint}}
\put(836.47,457.85){\usebox{\plotpoint}}
\put(854.59,467.95){\usebox{\plotpoint}}
\put(872.66,478.15){\usebox{\plotpoint}}
\put(890.81,488.21){\usebox{\plotpoint}}
\multiput(894,490)(18.314,9.767){0}{\usebox{\plotpoint}}
\put(909.09,498.05){\usebox{\plotpoint}}
\put(927.01,508.50){\usebox{\plotpoint}}
\put(945.34,518.21){\usebox{\plotpoint}}
\put(963.38,528.47){\usebox{\plotpoint}}
\put(981.55,538.50){\usebox{\plotpoint}}
\put(999.81,548.36){\usebox{\plotpoint}}
\multiput(1001,549)(17.798,10.679){0}{\usebox{\plotpoint}}
\put(1017.71,558.85){\usebox{\plotpoint}}
\put(1036.10,568.46){\usebox{\plotpoint}}
\put(1054.09,578.78){\usebox{\plotpoint}}
\put(1072.28,588.78){\usebox{\plotpoint}}
\put(1090.52,598.68){\usebox{\plotpoint}}
\multiput(1093,600)(17.798,10.679){0}{\usebox{\plotpoint}}
\put(1108.39,609.23){\usebox{\plotpoint}}
\put(1126.32,619.66){\usebox{\plotpoint}}
\put(1144.65,629.39){\usebox{\plotpoint}}
\put(1162.69,639.63){\usebox{\plotpoint}}
\put(1180.86,649.67){\usebox{\plotpoint}}
\put(1199.12,659.53){\usebox{\plotpoint}}
\multiput(1200,660)(17.798,10.679){0}{\usebox{\plotpoint}}
\put(1217.02,670.01){\usebox{\plotpoint}}
\put(1235.40,679.64){\usebox{\plotpoint}}
\put(1253.41,689.95){\usebox{\plotpoint}}
\put(1271.59,699.96){\usebox{\plotpoint}}
\put(1289.83,709.85){\usebox{\plotpoint}}
\multiput(1292,711)(17.798,10.679){0}{\usebox{\plotpoint}}
\put(1307.72,720.36){\usebox{\plotpoint}}
\put(1326.15,729.89){\usebox{\plotpoint}}
\put(1344.12,740.27){\usebox{\plotpoint}}
\put(1362.17,750.50){\usebox{\plotpoint}}
\put(1380.48,760.24){\usebox{\plotpoint}}
\put(1398.43,770.66){\usebox{\plotpoint}}
\multiput(1399,771)(18.314,9.767){0}{\usebox{\plotpoint}}
\put(1416.69,780.51){\usebox{\plotpoint}}
\put(1434.84,790.58){\usebox{\plotpoint}}
\put(1452.92,800.75){\usebox{\plotpoint}}
\put(1471.18,810.59){\usebox{\plotpoint}}
\put(1489.18,820.91){\usebox{\plotpoint}}
\multiput(1491,822)(18.314,9.767){0}{\usebox{\plotpoint}}
\put(1507.42,830.80){\usebox{\plotpoint}}
\put(1525.56,840.90){\usebox{\plotpoint}}
\put(1543.68,851.01){\usebox{\plotpoint}}
\put(1561.88,860.94){\usebox{\plotpoint}}
\put(1579.93,871.16){\usebox{\plotpoint}}
\multiput(1583,873)(18.314,9.767){0}{\usebox{\plotpoint}}
\put(1598.15,881.09){\usebox{\plotpoint}}
\put(1616.08,891.54){\usebox{\plotpoint}}
\put(1634.41,901.24){\usebox{\plotpoint}}
\put(1652.44,911.50){\usebox{\plotpoint}}
\put(1670.61,921.53){\usebox{\plotpoint}}
\put(1688.87,931.40){\usebox{\plotpoint}}
\multiput(1690,932)(17.798,10.679){0}{\usebox{\plotpoint}}
\put(1706.78,941.89){\usebox{\plotpoint}}
\put(1725.16,951.50){\usebox{\plotpoint}}
\put(1736,958){\usebox{\plotpoint}}
\sbox{\plotpoint}{\rule[-0.400pt]{0.800pt}{0.800pt}}%
\sbox{\plotpoint}{\rule[-0.500pt]{1.000pt}{1.000pt}}%
\put(220,131){\usebox{\plotpoint}}
\put(220.00,131.00){\usebox{\plotpoint}}
\put(238.85,139.68){\usebox{\plotpoint}}
\put(257.79,148.17){\usebox{\plotpoint}}
\put(276.32,157.50){\usebox{\plotpoint}}
\put(295.16,166.19){\usebox{\plotpoint}}
\multiput(297,167)(18.808,8.777){0}{\usebox{\plotpoint}}
\put(313.98,174.93){\usebox{\plotpoint}}
\put(332.72,183.86){\usebox{\plotpoint}}
\put(351.39,192.92){\usebox{\plotpoint}}
\put(370.20,201.69){\usebox{\plotpoint}}
\multiput(373,203)(18.808,8.777){0}{\usebox{\plotpoint}}
\put(388.99,210.50){\usebox{\plotpoint}}
\put(407.61,219.68){\usebox{\plotpoint}}
\put(426.41,228.46){\usebox{\plotpoint}}
\put(445.35,236.96){\usebox{\plotpoint}}
\put(463.83,246.38){\usebox{\plotpoint}}
\multiput(465,247)(18.808,8.777){0}{\usebox{\plotpoint}}
\put(482.64,255.15){\usebox{\plotpoint}}
\put(501.59,263.61){\usebox{\plotpoint}}
\put(520.15,272.88){\usebox{\plotpoint}}
\put(538.94,281.66){\usebox{\plotpoint}}
\multiput(542,283)(18.808,8.777){0}{\usebox{\plotpoint}}
\put(557.78,290.37){\usebox{\plotpoint}}
\put(576.53,299.27){\usebox{\plotpoint}}
\put(595.19,308.36){\usebox{\plotpoint}}
\put(614.00,317.13){\usebox{\plotpoint}}
\put(632.81,325.91){\usebox{\plotpoint}}
\multiput(633,326)(18.564,9.282){0}{\usebox{\plotpoint}}
\put(651.41,335.12){\usebox{\plotpoint}}
\put(670.21,343.90){\usebox{\plotpoint}}
\put(689.13,352.43){\usebox{\plotpoint}}
\put(707.66,361.75){\usebox{\plotpoint}}
\multiput(710,363)(18.808,8.777){0}{\usebox{\plotpoint}}
\put(726.42,370.62){\usebox{\plotpoint}}
\put(745.39,379.05){\usebox{\plotpoint}}
\put(763.98,388.26){\usebox{\plotpoint}}
\put(782.73,397.13){\usebox{\plotpoint}}
\put(801.58,405.81){\usebox{\plotpoint}}
\multiput(802,406)(18.808,8.777){0}{\usebox{\plotpoint}}
\put(820.35,414.67){\usebox{\plotpoint}}
\put(838.99,423.80){\usebox{\plotpoint}}
\put(857.80,432.57){\usebox{\plotpoint}}
\put(876.61,441.35){\usebox{\plotpoint}}
\multiput(878,442)(18.564,9.282){0}{\usebox{\plotpoint}}
\put(895.20,450.56){\usebox{\plotpoint}}
\put(914.01,459.34){\usebox{\plotpoint}}
\put(932.92,467.90){\usebox{\plotpoint}}
\put(951.80,476.51){\usebox{\plotpoint}}
\multiput(955,478)(18.314,9.767){0}{\usebox{\plotpoint}}
\put(970.21,486.09){\usebox{\plotpoint}}
\put(989.19,494.49){\usebox{\plotpoint}}
\put(1008.00,503.27){\usebox{\plotpoint}}
\put(1026.67,512.33){\usebox{\plotpoint}}
\put(1045.40,521.26){\usebox{\plotpoint}}
\multiput(1047,522)(18.808,8.777){0}{\usebox{\plotpoint}}
\put(1064.24,529.98){\usebox{\plotpoint}}
\put(1083.06,538.70){\usebox{\plotpoint}}
\put(1101.60,548.01){\usebox{\plotpoint}}
\put(1120.41,556.79){\usebox{\plotpoint}}
\multiput(1123,558)(19.015,8.319){0}{\usebox{\plotpoint}}
\put(1139.38,565.20){\usebox{\plotpoint}}
\put(1157.79,574.77){\usebox{\plotpoint}}
\put(1176.68,583.36){\usebox{\plotpoint}}
\put(1195.58,591.94){\usebox{\plotpoint}}
\put(1214.01,601.47){\usebox{\plotpoint}}
\multiput(1215,602)(19.015,8.319){0}{\usebox{\plotpoint}}
\put(1232.97,609.92){\usebox{\plotpoint}}
\put(1251.78,618.70){\usebox{\plotpoint}}
\put(1270.46,627.73){\usebox{\plotpoint}}
\put(1289.18,636.69){\usebox{\plotpoint}}
\multiput(1292,638)(18.808,8.777){0}{\usebox{\plotpoint}}
\put(1308.00,645.44){\usebox{\plotpoint}}
\put(1326.87,654.06){\usebox{\plotpoint}}
\put(1345.38,663.44){\usebox{\plotpoint}}
\put(1364.18,672.22){\usebox{\plotpoint}}
\put(1383.16,680.63){\usebox{\plotpoint}}
\multiput(1384,681)(18.314,9.767){0}{\usebox{\plotpoint}}
\put(1401.57,690.20){\usebox{\plotpoint}}
\put(1420.45,698.82){\usebox{\plotpoint}}
\put(1439.36,707.37){\usebox{\plotpoint}}
\put(1457.82,716.84){\usebox{\plotpoint}}
\multiput(1460,718)(19.015,8.319){0}{\usebox{\plotpoint}}
\put(1476.75,725.35){\usebox{\plotpoint}}
\put(1495.55,734.13){\usebox{\plotpoint}}
\put(1514.25,743.13){\usebox{\plotpoint}}
\put(1532.96,752.12){\usebox{\plotpoint}}
\put(1551.77,760.89){\usebox{\plotpoint}}
\multiput(1552,761)(19.015,8.319){0}{\usebox{\plotpoint}}
\put(1570.68,769.43){\usebox{\plotpoint}}
\put(1589.16,778.87){\usebox{\plotpoint}}
\put(1607.96,787.65){\usebox{\plotpoint}}
\put(1626.92,796.09){\usebox{\plotpoint}}
\multiput(1629,797)(18.314,9.767){0}{\usebox{\plotpoint}}
\put(1645.35,805.63){\usebox{\plotpoint}}
\put(1664.21,814.28){\usebox{\plotpoint}}
\put(1683.14,822.80){\usebox{\plotpoint}}
\put(1701.63,832.20){\usebox{\plotpoint}}
\put(1720.52,840.79){\usebox{\plotpoint}}
\multiput(1721,841)(18.808,8.777){0}{\usebox{\plotpoint}}
\put(1736,848){\usebox{\plotpoint}}
\end{picture}
\end{LARGE}
\caption{The bare quark mass dependence of the chiral order parameter for
the $\beta$ values of 4.125 (diamonds), 4.25 (squares), and 4.375 (triangles).
The error bars are smaller than the plotting symbols. The dotted lines are drawn
through the two lowest $am_q$ points at each value of $\beta$.} 
\label{pbpm}
\end{figure}

\vspace{0.3cm}
Although many of the gross features of lattice QCD with $N_f = 16$
thus seem to compare very
well with earlier studies at smaller values of $N_f$, we nevertheless
feel that a more precise characterization of the phase on the weak-coupling
side of our observed transition is called for. As some steps in this
direction, we have investigated the behavior of the chiral condensate
$\langle \bar{\psi}\psi \rangle$ and all the correlation functions as
functions of the quark mass $m_q$. 
The first question to answer is whether the weak-coupling
side of the phase transition really is chirally symmetric
in the limit of vanishing quark masses. To illustrate, we plot in
Figure \ref{pbpm} $\langle \bar{\psi}\psi\rangle$ as a function of $am_q$
for the different $\beta$-values quoted (at $\beta = 4.125$ taking values
from the weak-coupling phase). For all three $\beta$-values the linear 
extrapolation
of $\langle \bar{\psi}\psi\rangle$ from the smallest two quark masses to
$am_q=0$ is close to zero. (At $\beta=4.375$ actually all three values lie
on a straight line while, as discussed below, the $am_q=0.15$, $\beta=4.25$
is probably in the wrong phase.) 
This is our first evidence that the phase on the weak coupling
side is chirally symmetric. Another piece of evidence comes
from the fact that a quark-antiquark state with the pion quantum numbers
in this phase does not behave as pseudo-Goldstone boson. This follows from
Table 1, where we list a number of extracted mass values for the usual
(lattice) particle spectrum. The pion mass definitely does {\em not}
approach zero as $m_q$ is taken to zero, in this phase. It should be
mentioned here that the fits required to extract the pion and rho masses
on the chirally symmetric side were much more difficult to obtain, as
compared with the chirally broken side (whereas the simplest exponential
fit worked very well on the broken side, more terms were required on
the symmetric). This {\em could} perhaps be an indication the whole bound-state
particle concept may not be completely well-defined in the chirally
symmetric phase. 

\begin{figure}
\begin{LARGE}
\setlength{\unitlength}{0.240900pt}
\ifx\plotpoint\undefined\newsavebox{\plotpoint}\fi
\sbox{\plotpoint}{\rule[-0.200pt]{0.400pt}{0.400pt}}%
\begin{picture}(1800,1080)(0,0)
\font\gnuplot=cmr10 at 10pt
\gnuplot
\sbox{\plotpoint}{\rule[-0.200pt]{0.400pt}{0.400pt}}%
\put(220.0,113.0){\rule[-0.200pt]{0.400pt}{227.410pt}}
\put(220.0,176.0){\rule[-0.200pt]{4.818pt}{0.400pt}}
\put(198,176){\makebox(0,0)[r]{0.08}}
\put(1716.0,176.0){\rule[-0.200pt]{4.818pt}{0.400pt}}
\put(220.0,302.0){\rule[-0.200pt]{4.818pt}{0.400pt}}
\put(198,302){\makebox(0,0)[r]{0.1}}
\put(1716.0,302.0){\rule[-0.200pt]{4.818pt}{0.400pt}}
\put(220.0,428.0){\rule[-0.200pt]{4.818pt}{0.400pt}}
\put(198,428){\makebox(0,0)[r]{0.12}}
\put(1716.0,428.0){\rule[-0.200pt]{4.818pt}{0.400pt}}
\put(220.0,554.0){\rule[-0.200pt]{4.818pt}{0.400pt}}
\put(198,554){\makebox(0,0)[r]{0.14}}
\put(1716.0,554.0){\rule[-0.200pt]{4.818pt}{0.400pt}}
\put(220.0,679.0){\rule[-0.200pt]{4.818pt}{0.400pt}}
\put(198,679){\makebox(0,0)[r]{0.16}}
\put(1716.0,679.0){\rule[-0.200pt]{4.818pt}{0.400pt}}
\put(220.0,805.0){\rule[-0.200pt]{4.818pt}{0.400pt}}
\put(198,805){\makebox(0,0)[r]{0.18}}
\put(1716.0,805.0){\rule[-0.200pt]{4.818pt}{0.400pt}}
\put(220.0,931.0){\rule[-0.200pt]{4.818pt}{0.400pt}}
\put(198,931){\makebox(0,0)[r]{0.2}}
\put(1716.0,931.0){\rule[-0.200pt]{4.818pt}{0.400pt}}
\put(220.0,1057.0){\rule[-0.200pt]{4.818pt}{0.400pt}}
\put(198,1057){\makebox(0,0)[r]{0.22}}
\put(1716.0,1057.0){\rule[-0.200pt]{4.818pt}{0.400pt}}
\put(220.0,113.0){\rule[-0.200pt]{0.400pt}{4.818pt}}
\put(220,68){\makebox(0,0){0}}
\put(220.0,1037.0){\rule[-0.200pt]{0.400pt}{4.818pt}}
\put(388.0,113.0){\rule[-0.200pt]{0.400pt}{4.818pt}}
\put(388,68){\makebox(0,0){0.02}}
\put(388.0,1037.0){\rule[-0.200pt]{0.400pt}{4.818pt}}
\put(557.0,113.0){\rule[-0.200pt]{0.400pt}{4.818pt}}
\put(557,68){\makebox(0,0){0.04}}
\put(557.0,1037.0){\rule[-0.200pt]{0.400pt}{4.818pt}}
\put(725.0,113.0){\rule[-0.200pt]{0.400pt}{4.818pt}}
\put(725,68){\makebox(0,0){0.06}}
\put(725.0,1037.0){\rule[-0.200pt]{0.400pt}{4.818pt}}
\put(894.0,113.0){\rule[-0.200pt]{0.400pt}{4.818pt}}
\put(894,68){\makebox(0,0){0.08}}
\put(894.0,1037.0){\rule[-0.200pt]{0.400pt}{4.818pt}}
\put(1062.0,113.0){\rule[-0.200pt]{0.400pt}{4.818pt}}
\put(1062,68){\makebox(0,0){0.1}}
\put(1062.0,1037.0){\rule[-0.200pt]{0.400pt}{4.818pt}}
\put(1231.0,113.0){\rule[-0.200pt]{0.400pt}{4.818pt}}
\put(1231,68){\makebox(0,0){0.12}}
\put(1231.0,1037.0){\rule[-0.200pt]{0.400pt}{4.818pt}}
\put(1399.0,113.0){\rule[-0.200pt]{0.400pt}{4.818pt}}
\put(1399,68){\makebox(0,0){0.14}}
\put(1399.0,1037.0){\rule[-0.200pt]{0.400pt}{4.818pt}}
\put(1568.0,113.0){\rule[-0.200pt]{0.400pt}{4.818pt}}
\put(1568,68){\makebox(0,0){0.16}}
\put(1568.0,1037.0){\rule[-0.200pt]{0.400pt}{4.818pt}}
\put(1736.0,113.0){\rule[-0.200pt]{0.400pt}{4.818pt}}
\put(1736,68){\makebox(0,0){0.18}}
\put(1736.0,1037.0){\rule[-0.200pt]{0.400pt}{4.818pt}}
\put(220.0,113.0){\rule[-0.200pt]{365.204pt}{0.400pt}}
\put(1736.0,113.0){\rule[-0.200pt]{0.400pt}{227.410pt}}
\put(220.0,1057.0){\rule[-0.200pt]{365.204pt}{0.400pt}}
\put(45,585){\makebox(0,0){$\sigma$}}
\put(978,23){\makebox(0,0){$am_q$}}
\put(220.0,113.0){\rule[-0.200pt]{0.400pt}{227.410pt}}
\put(641,318){\raisebox{-.8pt}{\makebox(0,0){$\Diamond$}}}
\put(1062,786){\raisebox{-.8pt}{\makebox(0,0){$\Diamond$}}}
\put(641.0,240.0){\rule[-0.200pt]{0.400pt}{37.821pt}}
\put(631.0,240.0){\rule[-0.200pt]{4.818pt}{0.400pt}}
\put(631.0,397.0){\rule[-0.200pt]{4.818pt}{0.400pt}}
\put(1062.0,671.0){\rule[-0.200pt]{0.400pt}{55.407pt}}
\put(1052.0,671.0){\rule[-0.200pt]{4.818pt}{0.400pt}}
\put(1052.0,901.0){\rule[-0.200pt]{4.818pt}{0.400pt}}
\sbox{\plotpoint}{\rule[-0.400pt]{0.800pt}{0.800pt}}%
\put(431,246){\raisebox{-.8pt}{\makebox(0,0){$\Box$}}}
\put(641,250){\raisebox{-.8pt}{\makebox(0,0){$\Box$}}}
\put(1062,469){\raisebox{-.8pt}{\makebox(0,0){$\Box$}}}
\put(431.0,185.0){\rule[-0.400pt]{0.800pt}{29.390pt}}
\put(421.0,185.0){\rule[-0.400pt]{4.818pt}{0.800pt}}
\put(421.0,307.0){\rule[-0.400pt]{4.818pt}{0.800pt}}
\put(641.0,180.0){\rule[-0.400pt]{0.800pt}{33.726pt}}
\put(631.0,180.0){\rule[-0.400pt]{4.818pt}{0.800pt}}
\put(631.0,320.0){\rule[-0.400pt]{4.818pt}{0.800pt}}
\put(1062.0,312.0){\rule[-0.400pt]{0.800pt}{75.643pt}}
\put(1052.0,312.0){\rule[-0.400pt]{4.818pt}{0.800pt}}
\put(1052.0,626.0){\rule[-0.400pt]{4.818pt}{0.800pt}}
\sbox{\plotpoint}{\rule[-0.500pt]{1.000pt}{1.000pt}}%
\sbox{\plotpoint}{\rule[-0.600pt]{1.200pt}{1.200pt}}%
\put(641,237){\makebox(0,0){$\triangle$}}
\put(1062,362){\makebox(0,0){$\triangle$}}
\put(1483,761){\makebox(0,0){$\triangle$}}
\put(641.0,177.0){\rule[-0.600pt]{1.200pt}{28.908pt}}
\put(631.0,177.0){\rule[-0.600pt]{4.818pt}{1.200pt}}
\put(631.0,297.0){\rule[-0.600pt]{4.818pt}{1.200pt}}
\put(1062.0,314.0){\rule[-0.600pt]{1.200pt}{23.126pt}}
\put(1052.0,314.0){\rule[-0.600pt]{4.818pt}{1.200pt}}
\put(1052.0,410.0){\rule[-0.600pt]{4.818pt}{1.200pt}}
\put(1483.0,622.0){\rule[-0.600pt]{1.200pt}{66.970pt}}
\put(1473.0,622.0){\rule[-0.600pt]{4.818pt}{1.200pt}}
\put(1473.0,900.0){\rule[-0.600pt]{4.818pt}{1.200pt}}
\end{picture}
\end{LARGE}
\caption{The bare quark mass dependence of the string tension for
the $\beta$ values of 4.125 (squares), 4.25 (diamonds), and 4.375 (triangles).}
\label{stm}
\end{figure}
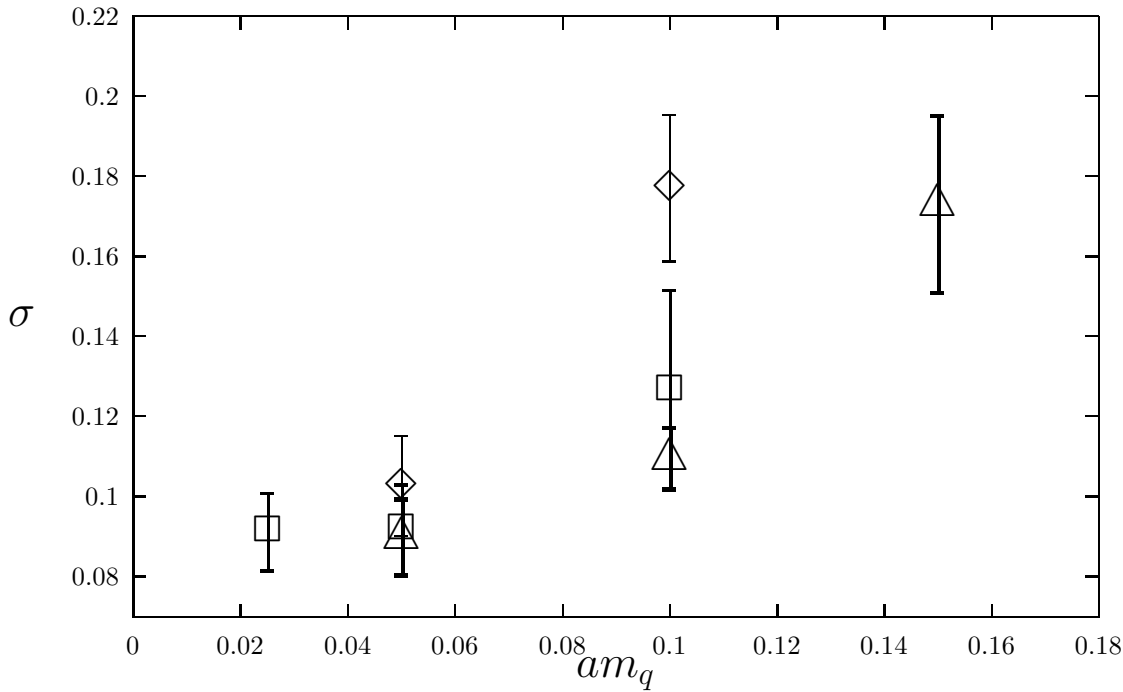

\begin{table}
\centerline{\begin{tabular}{|rrrrrrr|} \hline
{$am_q$} & {$\beta$} & {$\pi$} & {$\rho$} & {$\pi_2$} & {$\rho_2$} & {$N$} \\
\hline
{0.025} & {4.25} & {0.741} & {0.910} & {0.73} & {0.91} & {1.553} \\
{} & {} & {$\pm 0.002$} & {$ \pm 0.008$} & {$\pm 0.09$} & {$\pm 0.02$} 
& {$\pm 0.007$} \\
{0.05} & {4.125} & {0.749} & {0.977} & {0.92} & {1.003} & {1.605} \\
{} & {} & {$\pm 0.004$} & {$\pm 0.005$} & {$\pm 0.01$} & {$\pm 0.007$} &
{$\pm 0.007$} \\
{0.05} & {4.25} & {0.774} & {0.972} & {0.918} & {1.012} & {1.606} \\
{} & {} & {$\pm 0.002$} & {$\pm 0.003$} & {$\pm 0.006$} & {$\pm 0.005$} & 
{$\pm 0.004$} \\
{0.05} & {4.375} & {0.780} & {0.957} & {0.892} & {0.990} & {1.586} \\
{} & {} & {$\pm 0.025$} & {$\pm 0.004$} & {$\pm 0.008$} & {$\pm 0.003$}
& {$\pm 0.005$} \\
{0.1} & {4.0} & {0.754} & {1.57} & {******} & {1.74} & {2.37} \\
{} & {} & {$\pm 0.001$} & {$\pm 0.03$} & {******} & {$\pm 0.025$} & {$\pm 0.02$} \\
{0.1} & {4.125} & {0.784} & {1.47} & {1.61} & {1.64} & {2.27} \\
{} & {broken} & {$\pm 0.001$} & {$\pm 0.01$} & {$\pm 0.08$} & {$\pm 0.05$} &
{$\pm 0.02$} \\
{0.1} & {4.125} & {0.845} & {1.210} & {1.181} & {1.252} & {1.90} \\
{} & {symmetric} & {$\pm 0.002$} & {$\pm 0.008$} & {$\pm 0.006$} & {$\pm 0.006$}
& {$\pm 0.01$} \\
{0.1} & {4.25} & {0.871} & {1.123} & {1.085} & {1.152} & {1.788} \\
{} & {} & {$\pm 0.002$} & {$\pm 0.006$} & {$\pm 0.005$} & {$\pm 0.005$}
& {$\pm 0.005$} \\ 
{0.1} & {4.375} & {0.882} & {1.079} & {1.039} & {1.088} & {1.737} \\
{} & {} & {$\pm 0.002$} & {$\pm 0.004$} & {$\pm 0.005$} & {$\pm 0.007$} 
& {$\pm 0.003$} \\
{0.15} & {4.25} & {0.9323} & {1.562} & {1.676} & {1.74} & {2.51} \\
{} & {} & {$\pm 0.0006$} & {$\pm 0.006$} & {$\pm 0.004$} & {$\pm 0.02$}
& {$\pm 0.02$} \\
{0.15} & {4.375} & {0.992} & {1.314} & {1.289} & {1.348} & {1.989} \\
{} & {} & {$\pm 0.002$} & {$\pm 0.014$} & {$\pm 0.006$} & {$\pm 0.005$}
& {$\pm 0.0035$} \\ 
\hline
\end{tabular}}
\caption{Summary of the hadron masses.}
\label{mass_summary}
\end{table}
As $m_q$ increases, we expect that effectively fewer quark degrees of freedom
participate in screening, and that hence the observed phase transition 
should move towards
weaker coupling. Due to this, some (large) values of $am_q$ on the presumed
weak coupling side may actually correspond to the other phase. 
Consider figure \ref{stm}, where we plot
the ``string tension'' $\sigma$ extracted from the Creutz ratios 
(\ref{Creutz}). For the point $am_q = 0.15$ at $\beta = 4.25$ we find
$\sigma = 0.49(5)$ which is too large to appear in the figure.
The point clearly belongs to the strong-coupling phase, as could already
be suspected from the high value of $\langle \bar{\psi}\psi\rangle$ in
Figure \ref{pbpm}. Otherwise the
behavior of $\sigma$ as a function of $am_q$ is as expected: larger
$am_q$ corresponds to less screening, and hence, effectively, a larger
value of $\sigma$ from (\ref{Creutz}).

\vspace{0.3cm}
To understand these results in terms of the renormalization group, we
have also made an attempt to extract a rough estimate of the beta function
on the weak-coupling side of our phase transition. The rather crude
method we have employed consists in measuring the ratio $m_{\rho}/m_{\pi}$
at one coupling $\beta$ and for a given value of $am_q$. Choosing a
different coupling $\beta'$ (on the same side of the phase transition),
we vary $am_q$ until the ratio $m_{\rho}/m_{\pi}$ matches the value at
coupling $\beta$. On an infinite volume, this gives us the (scheme
dependent) scale change required to keep physics constant as we go from
$\beta$ to $\beta'$, and hence, indirectly, the beta function.
These results indicate that the lattice
spacing {\em grows} as we move towards weaker coupling, or, alternatively,
that the beta function in the usual convention is {\em positive} in this
region. This result is completely consistent with the two-loop perturbative
beta function which precisely, for $N_f = 16$, passes through a zero
close to the origin, and is positive beyond. If correct, the phase we
are investigating on the weak coupling side is a pure lattice artifact,
with no meaningful continuum limit (bounded as it is by a fixed point of
vanishing correlation length on one side, and a first order phase transition
on the other). The fact that this phase is strongly coupled but
chirally symmetric (we call it strongly coupled since, like in ordinary
QCD, the meson masses are much heavier than twice the quark mass)
is from this point of view just another lattice artifact.

\begin{figure}
\begin{LARGE}
\setlength{\unitlength}{0.240900pt}
\ifx\plotpoint\undefined\newsavebox{\plotpoint}\fi
\sbox{\plotpoint}{\rule[-0.200pt]{0.400pt}{0.400pt}}%
\begin{picture}(1800,1080)(0,0)
\font\gnuplot=cmr10 at 10pt
\gnuplot
\sbox{\plotpoint}{\rule[-0.200pt]{0.400pt}{0.400pt}}%
\put(220.0,113.0){\rule[-0.200pt]{0.400pt}{227.410pt}}
\put(220.0,113.0){\rule[-0.200pt]{4.818pt}{0.400pt}}
\put(198,113){\makebox(0,0)[r]{1.2}}
\put(1716.0,113.0){\rule[-0.200pt]{4.818pt}{0.400pt}}
\put(220.0,207.0){\rule[-0.200pt]{4.818pt}{0.400pt}}
\put(198,207){\makebox(0,0)[r]{1.25}}
\put(1716.0,207.0){\rule[-0.200pt]{4.818pt}{0.400pt}}
\put(220.0,302.0){\rule[-0.200pt]{4.818pt}{0.400pt}}
\put(198,302){\makebox(0,0)[r]{1.3}}
\put(1716.0,302.0){\rule[-0.200pt]{4.818pt}{0.400pt}}
\put(220.0,396.0){\rule[-0.200pt]{4.818pt}{0.400pt}}
\put(198,396){\makebox(0,0)[r]{1.35}}
\put(1716.0,396.0){\rule[-0.200pt]{4.818pt}{0.400pt}}
\put(220.0,491.0){\rule[-0.200pt]{4.818pt}{0.400pt}}
\put(198,491){\makebox(0,0)[r]{1.4}}
\put(1716.0,491.0){\rule[-0.200pt]{4.818pt}{0.400pt}}
\put(220.0,585.0){\rule[-0.200pt]{4.818pt}{0.400pt}}
\put(198,585){\makebox(0,0)[r]{1.45}}
\put(1716.0,585.0){\rule[-0.200pt]{4.818pt}{0.400pt}}
\put(220.0,679.0){\rule[-0.200pt]{4.818pt}{0.400pt}}
\put(198,679){\makebox(0,0)[r]{1.5}}
\put(1716.0,679.0){\rule[-0.200pt]{4.818pt}{0.400pt}}
\put(220.0,774.0){\rule[-0.200pt]{4.818pt}{0.400pt}}
\put(198,774){\makebox(0,0)[r]{1.55}}
\put(1716.0,774.0){\rule[-0.200pt]{4.818pt}{0.400pt}}
\put(220.0,868.0){\rule[-0.200pt]{4.818pt}{0.400pt}}
\put(198,868){\makebox(0,0)[r]{1.6}}
\put(1716.0,868.0){\rule[-0.200pt]{4.818pt}{0.400pt}}
\put(220.0,963.0){\rule[-0.200pt]{4.818pt}{0.400pt}}
\put(198,963){\makebox(0,0)[r]{1.65}}
\put(1716.0,963.0){\rule[-0.200pt]{4.818pt}{0.400pt}}
\put(220.0,1057.0){\rule[-0.200pt]{4.818pt}{0.400pt}}
\put(198,1057){\makebox(0,0)[r]{1.7}}
\put(1716.0,1057.0){\rule[-0.200pt]{4.818pt}{0.400pt}}
\put(220.0,113.0){\rule[-0.200pt]{0.400pt}{4.818pt}}
\put(220,68){\makebox(0,0){0}}
\put(220.0,1037.0){\rule[-0.200pt]{0.400pt}{4.818pt}}
\put(388.0,113.0){\rule[-0.200pt]{0.400pt}{4.818pt}}
\put(388,68){\makebox(0,0){0.02}}
\put(388.0,1037.0){\rule[-0.200pt]{0.400pt}{4.818pt}}
\put(557.0,113.0){\rule[-0.200pt]{0.400pt}{4.818pt}}
\put(557,68){\makebox(0,0){0.04}}
\put(557.0,1037.0){\rule[-0.200pt]{0.400pt}{4.818pt}}
\put(725.0,113.0){\rule[-0.200pt]{0.400pt}{4.818pt}}
\put(725,68){\makebox(0,0){0.06}}
\put(725.0,1037.0){\rule[-0.200pt]{0.400pt}{4.818pt}}
\put(894.0,113.0){\rule[-0.200pt]{0.400pt}{4.818pt}}
\put(894,68){\makebox(0,0){0.08}}
\put(894.0,1037.0){\rule[-0.200pt]{0.400pt}{4.818pt}}
\put(1062.0,113.0){\rule[-0.200pt]{0.400pt}{4.818pt}}
\put(1062,68){\makebox(0,0){0.1}}
\put(1062.0,1037.0){\rule[-0.200pt]{0.400pt}{4.818pt}}
\put(1231.0,113.0){\rule[-0.200pt]{0.400pt}{4.818pt}}
\put(1231,68){\makebox(0,0){0.12}}
\put(1231.0,1037.0){\rule[-0.200pt]{0.400pt}{4.818pt}}
\put(1399.0,113.0){\rule[-0.200pt]{0.400pt}{4.818pt}}
\put(1399,68){\makebox(0,0){0.14}}
\put(1399.0,1037.0){\rule[-0.200pt]{0.400pt}{4.818pt}}
\put(1568.0,113.0){\rule[-0.200pt]{0.400pt}{4.818pt}}
\put(1568,68){\makebox(0,0){0.16}}
\put(1568.0,1037.0){\rule[-0.200pt]{0.400pt}{4.818pt}}
\put(1736.0,113.0){\rule[-0.200pt]{0.400pt}{4.818pt}}
\put(1736,68){\makebox(0,0){0.18}}
\put(1736.0,1037.0){\rule[-0.200pt]{0.400pt}{4.818pt}}
\put(220.0,113.0){\rule[-0.200pt]{365.204pt}{0.400pt}}
\put(1736.0,113.0){\rule[-0.200pt]{0.400pt}{227.410pt}}
\put(220.0,1057.0){\rule[-0.200pt]{365.204pt}{0.400pt}}
\put(45,585){\makebox(0,0){$m_\rho\over m_\pi$}}
\put(978,23){\makebox(0,0){$am_q$}}
\put(220.0,113.0){\rule[-0.200pt]{0.400pt}{227.410pt}}
\put(641,310){\raisebox{-.8pt}{\makebox(0,0){$\Diamond$}}}
\put(1062,551){\raisebox{-.8pt}{\makebox(0,0){$\Diamond$}}}
\put(641.0,292.0){\rule[-0.200pt]{0.400pt}{8.672pt}}
\put(631.0,292.0){\rule[-0.200pt]{4.818pt}{0.400pt}}
\put(631.0,328.0){\rule[-0.200pt]{4.818pt}{0.400pt}}
\put(1062.0,532.0){\rule[-0.200pt]{0.400pt}{9.154pt}}
\put(1052.0,532.0){\rule[-0.200pt]{4.818pt}{0.400pt}}
\put(1052.0,570.0){\rule[-0.200pt]{4.818pt}{0.400pt}}
\sbox{\plotpoint}{\rule[-0.400pt]{0.800pt}{0.800pt}}%
\put(431,166){\raisebox{-.8pt}{\makebox(0,0){$\Box$}}}
\put(641,218){\raisebox{-.8pt}{\makebox(0,0){$\Box$}}}
\put(1062,282){\raisebox{-.8pt}{\makebox(0,0){$\Box$}}}
\put(1483,1011){\raisebox{-.8pt}{\makebox(0,0){$\Box$}}}
\put(431.0,145.0){\rule[-0.400pt]{0.800pt}{10.118pt}}
\put(421.0,145.0){\rule[-0.400pt]{4.818pt}{0.800pt}}
\put(421.0,187.0){\rule[-0.400pt]{4.818pt}{0.800pt}}
\put(641.0,209.0){\rule[-0.400pt]{0.800pt}{4.577pt}}
\put(631.0,209.0){\rule[-0.400pt]{4.818pt}{0.800pt}}
\put(631.0,228.0){\rule[-0.400pt]{4.818pt}{0.800pt}}
\put(1062.0,267.0){\rule[-0.400pt]{0.800pt}{6.986pt}}
\put(1052.0,267.0){\rule[-0.400pt]{4.818pt}{0.800pt}}
\put(1052.0,296.0){\rule[-0.400pt]{4.818pt}{0.800pt}}
\put(1483.0,998.0){\rule[-0.400pt]{0.800pt}{6.022pt}}
\put(1473.0,998.0){\rule[-0.400pt]{4.818pt}{0.800pt}}
\put(1473.0,1023.0){\rule[-0.400pt]{4.818pt}{0.800pt}}
\sbox{\plotpoint}{\rule[-0.500pt]{1.000pt}{1.000pt}}%
\sbox{\plotpoint}{\rule[-0.600pt]{1.200pt}{1.200pt}}%
\put(641,164){\makebox(0,0){$\triangle$}}
\put(1062,157){\makebox(0,0){$\triangle$}}
\put(1483,348){\makebox(0,0){$\triangle$}}
\put(641.0,152.0){\rule[-0.600pt]{1.200pt}{5.782pt}}
\put(631.0,152.0){\rule[-0.600pt]{4.818pt}{1.200pt}}
\put(631.0,176.0){\rule[-0.600pt]{4.818pt}{1.200pt}}
\put(1062.0,148.0){\rule[-0.600pt]{1.200pt}{4.336pt}}
\put(1052.0,148.0){\rule[-0.600pt]{4.818pt}{1.200pt}}
\put(1052.0,166.0){\rule[-0.600pt]{4.818pt}{1.200pt}}
\put(1483.0,321.0){\rule[-0.600pt]{1.200pt}{13.009pt}}
\put(1473.0,321.0){\rule[-0.600pt]{4.818pt}{1.200pt}}
\put(1473.0,375.0){\rule[-0.600pt]{4.818pt}{1.200pt}}
\end{picture}
\end{LARGE}
\caption{The bare quark mass dependence of the $\rho$ to pion mass ratio for
the $\beta$ values of 4.125 (diamonds), 4.25 (squares), and 4.375 (triangles).}
\label{rhopi}
\end{figure}
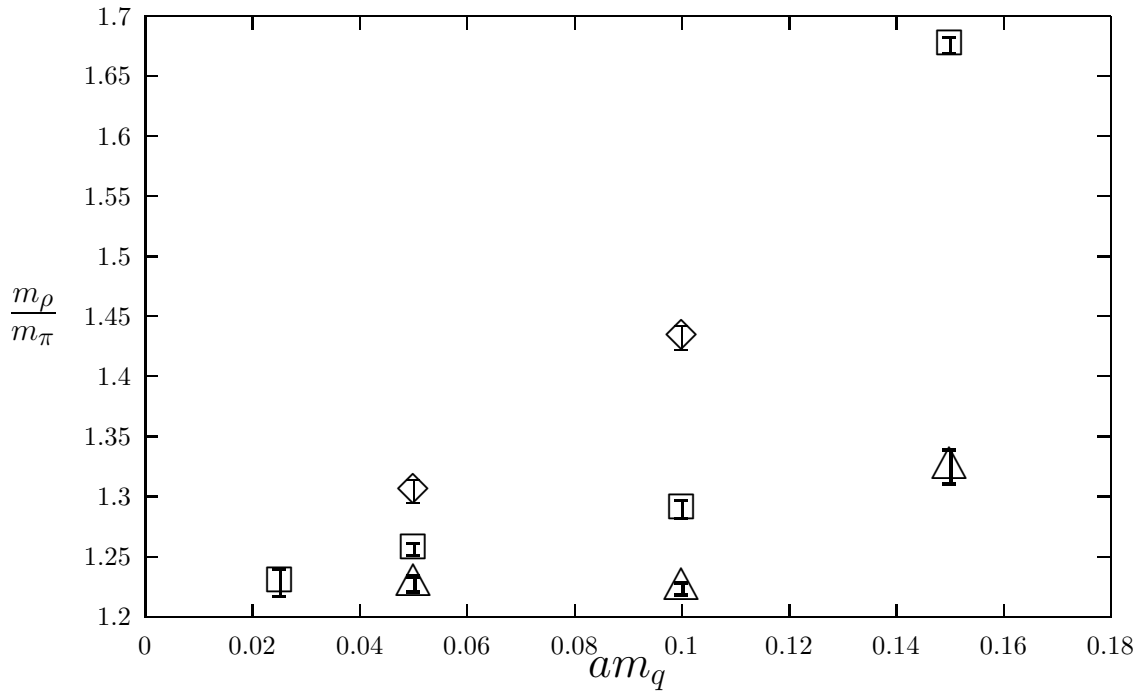

If this picture were the whole story, we should presently be able to
pursue directly our goal: to find the strict upper limit of light
quark flavors $N_f$ which is still compatible with chiral symmetry breaking
and confinement.\footnote{The weak-coupling phase we would see directly
in such lattice simulations would correspond to the branch of the
beta function beyond the infrared fixed point, with no continuum limit. 
But already
the mere existence of an infrared fixed point at relatively weak
coupling is presumable enough to prevent chiral symmetry breaking, and
almost certainly strict confinement, also on the (physical) branch connected 
to the origin.} 
Namely, we could lower $N_f$ and follow the weak-coupling
chirally symmetric phase. Measuring the beta function, we could find the
precise point (as a function of $N_f$) where the infrared fixed point
disappears. However, there are a number of puzzling results, as well as
a more fundamental problem, that prevent
us from immediately doing so. For example, if we compare with the very 
precise study
of the $N_f=8$ theory in ref. \cite{Christ}, we note that both the phase
transition and the weak-coupling side of it look remarkably similar to
what we find in the $N_f = 16$ theory. The apparently vanishing chiral
condensate, the individual scaling of the mass spectrum in this phase:
both are very similar in the $N_f = 8$ and the $N_f = 16$ theories.
Naively, this would indicate that already a number $N_f = 8$ of quark
flavors is incompatible with chiral symmetry breaking (and confinement).
But in detail the results differ. For example, if we use the mass
measurements in ref. \cite{Christ} to find the analogous beta function
(by varying $am_q$, while keeping $m_{\rho}/m_{\pi}$ fixed, using 
mild extrapolations of their data), we find that
it has precisely the opposite sign. Of course, the ``beta function'' we
can extract in this manner is by no means universal, though taking the
nucleon instead of the $\rho$ gives qualitatively the same result.
While it is a crude approximation to perform the matching in the same
space-time volumes, the scale change between, for example $\beta=4.125$,
$am_q=0.05$ and $\beta=4.25$, $am_q=0.1$ is only about 1.14 from the
$\rho$ mass, and hence the finite volume effects should not be too
different and largely cancel. There seems to be a large anomalous scale
factor for the quark mass. If the
present ``scheme'' is sensible, we would not expect it to be able to 
introduce spurious zeros in the beta function so defined. Then the
difference in sign of the beta functions for $N_f = 8$ and $N_f = 16$,
which agrees with the perturbative prediction once, for $N_f = 16$,
we are on the strong coupling side of the perturbative critical point,
eq.~(\ref{su3}), has to be taken seriously. 
We are presently investigating these scenarios in more detail.
Another more fundamental
problem concerns the non-universality of the beta function far away
from fixed points of diverging correlation lengths. The precise location
of an infrared fixed point at some intermediate coupling, and even the 
whole meaning of it, is scheme dependent. In lattice language, fixed
points may develop which are to be viewed as no more than lattice
artifacts. Therefore, even in principle it is a highly non-trivial
task to pin-point the precise number of flavors at which confinement
is lost.

\vspace{0.3cm}
Finally we wish to mention that
contrary to the arguments of ref. \cite{Iwasaki}, we see no 
reasons or numerical indications
whatsoever for sensitivity to $N_f$ on the extreme strong-coupling side.
At very strong coupling there is no special significance to attach to the
numbers $N_f = 16$ and $N_f = 17$, in contrast to the situation at
weak coupling. Certainly, for $N_f \geq 17$ we do not expect to be able
to define a non-trivial continuum theory anywhere. 
But at strong coupling neither
this value, nor values such as $N_f = 6$ or $N_f = 7$ \cite{Iwasaki}, 
play any special r\^{o}les. The reason why the two pictures are
totally compatible, is precisely to be found in the existence of a
discontinuous bulk phase transition separating the two regimes. 
At least for staggered lattice fermions we thus find no support for the kind 
of scenario suggested, for Wilson fermions, in ref. \cite{Iwasaki}. A more
interesting question concerns the fate of the bulk phase 
transition as $N_f$ is lowered, and, in particular, the nature of the
phases on the weak-coupling side. Also this is presently under study. 

\vspace{0.3cm}
\noindent
{\sc Acknowledgements:} This work has been supported in part by the Danish
National Research Council (SNF) and by the US Department of Energy
under grants \#~DE-FG05-85ER250000 and \#~DE-FG05-95ER40979.
This work was in part based on the MILC collaboration's public lattice
gauge theory code \cite{MILC_code}.
Computing facilities for the simulations were provided by UNI-C and by SCRI.

\vspace{0.5cm}


\end{document}